\documentclass[journal,twocolumn]{IEEEtran}
\usepackage{ifpdf}
\usepackage{cite}
\usepackage{graphicx} 
\usepackage{amsmath,amsfonts,bm,amssymb,mathtools}
\usepackage{mathalfa}
\usepackage[thinc]{esdiff}
\usepackage{caption}
\usepackage{balance}
\usepackage{booktabs}
\usepackage{tikz}
\usepackage{subcaption}
\usepackage[font=footnotesize]{caption}
\usepackage{algorithm,algorithmicx,algpseudocode}

\newcommand{\lr}[1]{\langle #1 \rangle}
\newcommand{\blr}[1]{\big\langle #1 \big\rangle}
\newcommand{\bblr}[1]{\bigg\langle #1 \bigg\rangle}
\newcommand{\Blr}[1]{\Big\langle #1 \Big\rangle}

\newcommand{\ji}{{\sf j}}

\newtheorem{theorem}{Theorem}

\newtheorem{corollary}{Corollary}

\newcommand{\bs}[1]{\boldsymbol{#1}}

\newcommand{\mc}[1]{\mathcal{#1}}

\newcommand{\mb}[1]{\mathbf{#1}}
\newcommand{\mr}[1]{\mathrm{#1}}
\newcommand{\tr}{\mathrm{Tr}}

\newcommand{\e}{\mathrm{e}}

\newcommand{\p}{\mathrm{p}}

\newcommand{\da}{\mathrm{d}}

\newcommand{\E}{\mathbb{E}}
\newcommand{\T}{\mathrm{T}}

\title{Joint Channel Estimation, Data Detection, and Phase Noise Tracking in Massive MIMO Systems}
\author{Toan-Van Nguyen and Duy H. N. Nguyen
\thanks{The authors are with the Department of Electrical and Computer Engineering, San Diego State University, San Diego, CA 92182 USA (emails:\{tnguyen58, duy.nguyen\}@sdsu.edu).}}

\begin{document}
\maketitle
\begin{abstract}
    Communication over multiple-input multiple-output (MIMO) systems is degraded by phase noise (PN) arising from imperfect local oscillators, with PN becoming more severe at higher carrier frequencies. Existing methods either rely on linearized Kalman-filter approaches or apply variational Bayes (VB) only to individual stages, such as data detection under a static phase prior, or channel estimation alone, leaving no unified framework that tracks an evolving PN process consistently across pilot and data transmission. This paper develops such a framework for uplink massive MIMO. Specifically, we first propose VB-Tracking, which employs von Mises priors and coordinate-ascent VB inference to jointly estimate the channel and track the transmit and receive PN trajectories during the pilot phase without linearization. We then propose VB-Detecting for joint data detection and phase tracking by absorbing the transmit PN into a composite transmit variable. Using a sequentially propagated von Mises predictive prior, the composite variable admits an exact closed-form posterior. Finally, we propose VB-Refining, which periodically re-estimates the channel during data transmission using the soft posterior moments of reliably detected transmit symbols as virtual pilots. We show that VB-Detecting scales linearly in the number of users, receive antennas, constellation size, and iterations. Simulations against EKF-EP, EKF-ZF, and JED baselines show that VB-Tracking, VB-Detecting, and VB-Refining together achieve lower channel NMSE and symbol error rate, with the performance gap widening at high SNR and under severe PN.
\end{abstract}
\begin{IEEEkeywords}
	Channel estimation, data detection, MIMO, phase noise, variational Bayes.
\end{IEEEkeywords}
\section{Introduction}
    Coherent communication over wireless multiple-input multiple-output (MIMO) links relies on stable phase references at both the transmitter and the receiver. However, in practice, every radio-frequency (RF) chain is driven by an imperfect free-running local oscillator, whose short-term frequency instability manifests as a randomly time-varying phase, often modeled as a Wiener (Brownian-motion) process, superimposed on the transmitted and received signals \cite{chang2024modeling,nasir2013phase}. This phase noise (PN) rotates the received constellation symbol-by-symbol and, in multicarrier systems, spreads energy across subcarriers as inter-carrier interference, degrading detection performance even when the channel itself is perfectly known \cite{mehrpouyan2012joint}. The impairment is far from a corner case for next-generation systems. As carriers move into millimeter-wave (mmWave) and sub-THz bands to access wider bandwidths, the PN power of the oscillator increases sharply with carrier frequency, while practical phase-locked loop designs at those frequencies can suppress it less effectively; thus, PN becomes a primary bottleneck rather than a secondary effect \cite{neshaastegaran2020effect,bicais2021design}. Massive MIMO further compounds the problem because it employs many RF chains, whether due to per-antenna oscillators or a modular array of tiles. Each chain is synchronized to a common reference but free-running relative to the others, which introduces distinct PN trajectories across the transmit and receive chains. As a result, the number of latent phase processes that must be tracked can scale with the number of antennas and users rather than remaining fixed \cite{nasir2013phase,rasekh2021phase,zheng2020joint}.  

    A substantial body of research has focused on PN mitigation at the receiver, which can be broadly grouped into three categories. First, extended Kalman filter (EKF) approaches use pilots and estimated data to recursively track Wiener PN trajectories, either jointly across the array \cite{mehrpouyan2012joint} or with independent per-oscillator states with forward–backward smoothing \cite{reggiani2018extended}. EKF-based tracking has also been applied to rapid PN variations in mmWave full-duplex systems \cite{koohian2019joint} and pilot-based PN tracking \cite{hajiabdolrahim2020extended}. Note that although the Kalman filter is an optimal minimum mean-square error (MMSE) estimator for linear-Gaussian state-space models \cite{kay1993fundamentals}, this optimality does not extend to the EKF. Since the EKF relies on a first-order linearization of the nonlinear observation model around the current state estimate, its tracking performance depends on the accuracy of this local linearization \cite{kay1993fundamentals}.
    Second, expectation-maximization (EM) methods alternate between estimating PN and transmitted symbols, using steepest-descent updates \cite{tarable2024compensation} or variational-EM with factorized Gaussian posteriors \cite{cheng2019compensation}. These methods are sensitive to initialization and local optima. Moreover, variational-EM formulations based on factorized posteriors neglect posterior correlations among the channel, PN, and transmitted symbols. Third, factor-graph message passing methods represent the PN posterior with a Tikhonov (von Mises) mixture and combine belief propagation, mean-field, and expectation propagation (EP) to jointly perform PN estimation, equalization, and decoding \cite{shayovitz2015message,wang2016bp}. EP-based detection has also been studied under strong Wiener PN \cite{conti2025application}. Other approaches balance complexity via approximate maximum-likelihood detection that bypasses PN estimation entirely, such as self-interference whitening \cite{combes2017approximate}, or used reliably detected symbols as virtual pilots to combat channel aging under time-varying PN \cite{wang2025iterative}.

    Variational Bayes (VB) inference provides an alternative framework that can yield closed-form coordinate-ascent updates in place of the linearization or moment-matching approximations used by EKF- and EP-based methods. Early applications of VB inference to PN-impaired orthogonal frequency-division multiplexing (OFDM) systems jointly estimated the channel and the full PN sequence by minimizing a variational free energy, showing EM to be a special case of the resulting estimator \cite{lin2006joint,lin2007variational}. A related line applied variational bounding to iterative detection/decoding under both constant and dynamic phase-uncertainty models \cite{nissila2009adaptive}. For MIMO systems, a VB framework built on free-energy maximization was proposed for joint PN estimation and symbol detection \cite{krishnan2012variational}, and closed-form VB updates using Tikhonov canonical distributions were shown to outperform smoother-detector and sum-product alternatives under quasi-static fading with Wiener PN \cite{krishnan2015algorithms}. Recently, generalized-expectation-consistent approximate Bayesian inference has been used for low-complexity symbol detection under static PNs in massive MIMO \cite{yang2019symbol}, and VB inference itself has been shown to yield strong multiuser massive MIMO detectors in the PN-free setting by treating the noise precision as latent variables \cite{nguyen2022variational}. A VB framework has extended joint channel estimation and data detection (JED) to few-bit massive MIMO receivers with an unknown noise variance \cite{nguyen2024variational,nguyen2026mimo}. Based on this VB-for-detection perspective, our own prior work derived an exact closed-form von Mises posterior for joint PN estimation and data detection, absorbing the transmit PN into the transmitted signal, but under a static PN instead of time-varying PN models \cite{nguyen2026variational}.

\subsection{Motivations and Contributions}
These aforementioned works reveal a common trade-off between accuracy and tractability. EKF-based trackers \cite{mehrpouyan2012joint,reggiani2018extended,koohian2019joint,hajiabdolrahim2020extended} rely on first-order linearization of the PN observation model, which loses accuracy under large phase innovations and becomes harder to sustain as the number of independent oscillators grows with array size. EM and variational-EM schemes \cite{tarable2024compensation,cheng2019compensation} track only point estimates of the PN, discarding posterior uncertainty that could inform detection reliability. Message-passing and Tikhonov-mixture receivers \cite{shayovitz2015message,wang2016bp,conti2025application} retain this uncertainty but require control of the mixture order through KL-divergence-based pruning and merging of components, as well as moment-matching approximations to project propagated messages back onto single Tikhonov/von Mises distributions.
These operations add computational complexity that grows with constellation size and the number of users, limiting the scalability of such approaches to massive MIMO systems. Meanwhile, existing VB-based PN methods are largely divided at the pilot-data boundary. Existing VB-based studies have primarily addressed either channel estimation under PN  \cite{lin2006joint,lin2007variational} or PN-aware data detection under static or quasi-static phase models \cite{nguyen2026variational,nissila2009adaptive,krishnan2012variational,krishnan2015algorithms}. Consequently, they do not provide a unified VB pipeline that carries an evolving Wiener-PN posterior consistently from the pilot phase through data detection and channel refinement. A JED-type VB framework was developed for massive MIMO \cite{nguyen2024variational}; however, it either omits PN or absorbs it directly into the channel estimate, thereby discarding the underlying phase-evolution structure and, as our simulations show, resulting in an SER floor under time-varying PN. Virtual-pilot channel refinement using detected data symbols has been explored; it has so far been paired only with EKF-based PN tracking \cite{wang2025iterative}, inheriting the same linearization sensitivity. No prior work carries a consistent, evolving Wiener PN posterior from the pilot phase through data detection and refinement. To close this gap, our contributions are as follows:
\begin{itemize}
    \item We first develop a \textbf{VB-Tracking} approach for joint channel estimation and phase noise tracking during the pilot transmission in uplink massive MIMO systems. Employing von Mises distributions to model the PN innovations and treating the noise precision as a latent variable, we derive closed-form coordinate-ascent variational inference (CAVI) updates for the channel and PN trajectories jointly, avoiding the EKF-style linearization.
    \item We then develop a \textbf{VB-Detecting} approach for joint data detection and PN tracking during data transmission. We absorb the transmit PN into the transmitted signal and treat the composite variable as the inference target. Unlike \cite{nguyen2026variational} based on a static PN prior, here the phase evolves via a Wiener recursion; thus the prior is instead a sequentially propagated von Mises predictive prior. Combined with a discrete data prior, it yields an exact conjugate posterior in closed form, presented as a finite mixture of von Mises distributions on concentric circles indexed by the constellation points. 
    \item We develop a \textbf{VB-Refining} approach for periodic channel re-estimation during data transmission, in which the soft posterior moments of reliably detected symbols are exploited as virtual pilots for periodic channel re-estimation, thereby mitigating channel aging during long-frame transmissions.
    \item A complexity analysis and extensive simulation are presented. We show that VB-Detecting scales linearly in the number of users, receive antennas, constellation size, and iterations. Simulations against EKF-EP, EKF-ZF, and JED baselines confirm that VB-Tracking, VB-Detecting, and VB-Refining together achieve substantially lower channel normalized mean squared error (NMSE) and SER, particularly at high SNR and under severe PN, where the gap over baselines widens.
	\end{itemize}

	\underline{\textit {Notation}}: Boldface lowercase and boldface uppercase variables denote vectors and matrices, respectively. The $L_2$-norm and the absolute value are indicated by $\|\cdot\|$ and $|\cdot|$, respectively. Real and imaginary parts are denoted by $\Re\{\cdot\}$ and $\Im\{\cdot\}$, respectively, with $\ji =\sqrt{-1}$. The PDF of a complex multivariate Gaussian random vector $\mathbf{x} \in \mathbb{C}^K$ with mean $\boldsymbol{\mu}$ and covariance matrix $\boldsymbol{\Sigma}$ is given by $\mathcal{CN}(\mathbf{x};\boldsymbol{\mu},\boldsymbol{\Sigma}) = \frac{1}{\pi^K|\boldsymbol{\Sigma}|}\exp\!\left(-(\mathbf{x}-\boldsymbol{\mu})^H\boldsymbol{\Sigma}^{-1}(\mathbf{x}-\boldsymbol{\mu})\right)$  where $\mathcal{CN}(\boldsymbol{\mu},\boldsymbol{\Sigma})$ is used as shorthand to denote the distribution itself. Likewise, the PDF of a von Mises random variable $x$ with mean $\mu$ and concentration $\kappa$ is given by $\mc{VM}(x;\mu,\kappa) = \frac{1}{2\pi I_0(\kappa)}\exp\bigl(\kappa \cos(x-\mu)\bigr)$ where $\mc{VM}(\mu,\kappa)$ is used as shorthand to denote the distribution itself. Equivalently, defining the complex concentration parameter $\eta \triangleq \kappa e^{\ji\mu} \in \mathbb{C}$ such that $\kappa = |\eta|$ and $\mu = \angle\eta$, the PDF and distribution are written as $\mathcal{VM}(x;\eta)$ and $\mathcal{VM}(\eta)$, respectively. The identity matrix is denoted by $\mb{I}$, the trace operator by $\tr$($\cdot$), and the expectation operator by $\mathbb{E}\{\cdot\}$. The transpose, complex conjugate, and complex conjugate transpose operators are denoted by $(\cdot)^T$, $(\cdot)^*$, and $(\cdot)^H$, respectively. The symbols $\propto$ and $\sim$ stand for ``is proportional to'' and ``distributed according to'', respectively. The bracket $\lr{\cdot}$ represents expectation with respect to all latent variables except the one under consideration. 
	
	\section{System Model}
	\subsection{System Model}
	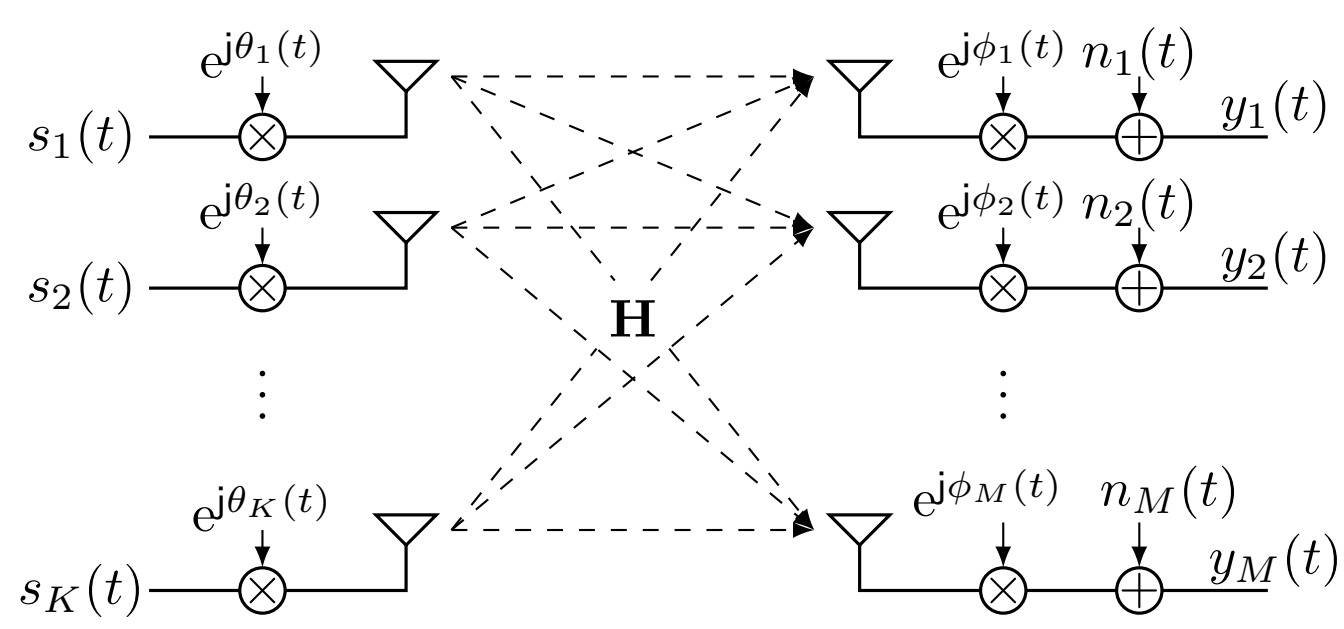
\begin{figure}
		\centering
		\begin{tikzpicture}
			\node at (-0.65,4) {$s_{1}(t)$};
			\draw [semithick] (-0.2,4) to (0.4,4);
			\draw [semithick] (0.55,4) circle [radius=0.15] node {$\times$};
			\draw [-latex] (0.55,4.4) to (0.55,4.15);
			\node at (0.555,4.55) {$\e^{\ji\theta_1(t)}$};
			\draw [semithick] (0.7,4) to (1.5,4) to (1.5,4.3) to (1.3,4.5) to (1.7,4.5) to (1.5,4.3);
			
			\node at (-0.65,3) {$s_{2}(t)$};
			\draw [semithick] (-0.2,3) to (0.4,3);
			\draw [semithick] (0.55,3) circle [radius=0.15] node {$\times$};
			\draw [-latex] (0.55,3.4) to (0.55,3.15);
			\node at (0.55,3.55) {$\e^{\ji\theta_2(t)}$};
			\draw [semithick] (0.7,3) to (1.5,3) to (1.5,3.3) to (1.3,3.5) to (1.7,3.5) to (1.5,3.3);
			
			\node at (-0.65,1) {$s_{K}(t)$};
			\draw [semithick] (-0.2,1) to (0.4,1);
			\draw [semithick] (0.55,1) circle [radius=0.15] node {$\times$};
			\draw [-latex] (0.55,1.4) to (0.55,1.15);
			\node at (0.55,1.55) {$\e^{\ji\theta_K(t)}$};
			\draw [semithick] (0.7,1) to (1.5,1) to (1.5,1.3) to (1.3,1.5) to (1.7,1.5) to (1.5,1.3);
			
			\node at (0.55,2.4) {$\vdots$};
			
			\node at (3,2.8) {$\mb{H}$};
			
			\draw [semithick] (5.3,4) to (4.5,4) to (4.5,4.3) to (4.3,4.5) to (4.7,4.5) to (4.5,4.3);
			\draw [semithick] (5.45,4) circle [radius=0.15] node {$\times$};
			\draw [-latex] (5.45,4.4) to (5.45,4.15);
			\node at (5.45,4.55) {$\e^{\ji\phi_1(t)}$};
			\draw [semithick] (5.6,4) to (6.2,4);
			\draw [semithick] (6.35,4) circle [radius=0.15] node {$+$};
			\draw [-latex] (6.35,4.4) to (6.35,4.15);
			\node at (6.35,4.55) {$n_1(t)$};
			\draw [semithick] (6.5,4) to (7.2,4);
			\node at (7.25,4.2) {$y_{1}(t)$};
			
			\draw [semithick] (5.3,3) to (4.5,3) to (4.5,3.3) to (4.3,3.5) to (4.7,3.5) to (4.5,3.3);
			\draw [semithick] (5.45,3) circle [radius=0.15] node {$\times$};
			\draw [-latex] (5.45,3.4) to (5.45,3.15);
			\node at (5.45,3.55) {$\e^{\ji\phi_2(t)}$};
			\draw [semithick] (5.6,3) to (6.2,3);
			\draw [semithick] (6.35,3) circle [radius=0.15] node {$+$};
			\draw [-latex] (6.35,3.4) to (6.35,3.15);
			\node at (6.35,3.55) {$n_2(t)$};
			\draw [semithick] (6.5,3) to (7.2,3);
			\node at (7.25,3.2) {$y_{2}(t)$};
			
			\draw [semithick] (5.3,1) to (4.5,1) to (4.5,1.3) to (4.3,1.5) to (4.7,1.5) to (4.5,1.3);
			\draw [semithick] (5.45,1) circle [radius=0.15] node {$\times$};
			\draw [-latex] (5.45,1.45) to (5.45,1.15);
			\node at (5.35,1.65) {$\e^{\ji\phi_M(t)}$};
			\draw [semithick] (5.6,1) to (6.2,1);
			\draw [semithick] (6.35,1) circle [radius=0.15] node {$+$};
			\draw [-latex] (6.35,1.45) to (6.35,1.15);
			\node at (6.55,1.65) {$n_M(t)$};
			\draw [semithick] (6.5,1) to (7.2,1);
			\node at (7.25,1.2) {$y_{M}(t)$};	
			
			\node at (5.45,2.4) {$\vdots$};
			
			\draw [-latex,dashed] (1.8,4.4) to (4.2,4.4);
			\draw [-latex,dashed] (1.8,4.4) to (4.2,3.4);
			\draw [dashed] (1.8,4.4) to (2.88,3.05);
			\draw [-latex,dashed] (3.24,2.6) to (4.2,1.4);
			\draw [-latex,dashed] (1.8,3.4) to (4.2,4.4);
			\draw [-latex,dashed] (1.8,3.4) to (4.2,3.4);
			\draw [-latex,dashed] (1.8,3.4) to (4.2,1.4);
			\draw [-latex,dashed] (1.8,1.4) to (4.2,1.4);
			\draw [-latex,dashed] (1.8,1.4) to (4.2,3.4);	
			\draw [dashed] (1.8,1.4) to (2.76,2.6);
			\draw [-latex,dashed] (3.12,3.05) to (4.2,4.4);
		\end{tikzpicture}
		\caption{Diagram of a massive MIMO system with $K$ transmit antennas and $M$ receive antennas. At time $t$, the transmitted signal at user-$i$ is corrupted by the PN factor $\e^{\ji\theta_i(t)}$, whereas the received signal at antenna-$m$ is corrupted by the PN factor $\e^{\ji\theta_m(t)}$ and the noise $n_m(t)$.}
		\label{figure-model}
	\end{figure}
We consider an uplink massive MIMO system as shown in Fig. \ref{figure-model}, where a base station (BS) with $M$ antennas is deployed to serve $K$ users under phase noise.
We consider a block fading channel consisting of a pilot transmission phase over $T_\p$ time slots followed by a data transmission phase over $T_\da$ time slots.
	The received signal at time slot $t$ is given by
	\begin{align} \label{eq:y}
		{\mb{y}}(t)&= \bs{\Phi}(t)\mb{H}\bs{\Theta}(t) {\mb{s}}(t)+ {\mb{n}}(t),
	\end{align}
	where $\bs{\Theta}(t)= \mr{diag}\big(\e^{\ji \theta_{1}(t)},\ldots,\e^{\ji \theta_{K}(t)}\big)$ and $\bs{\Phi}(t) = \mr{diag}\big(\e^{\ji \phi_1(t)},\ldots,\e^{\ji \phi_{M}(t)}\big)$ are $K\times K$ and $M\times M$ matrices capturing the phase noises at $K$ users and the BS, respectively, $\mb{H} = [\mb{h}_1,\ldots,\mb{h}_K] \in \mathbb{C}^{M \times K}$ is the uplink channel matrix, with $\mb{h}_i \in \mathbb{C}^{M\times 1}$, and $\mb{s}(t) = [s_{1}(t),\ldots,s_{K}(t)]^\T \in \mc{A}^{K}$ is the transmitted vector. 
	The prior distribution of $s_{i}(t)$ is given by
	\begin{align}
		p(s_{i}(t)) = \sum_{a \in \mc{A}}p_a\delta(s_{i}(t) - a),
	\end{align}
	where $p_a$ is the known prior probability of the constellation point $a$ and $\delta(s_{i}(t)-a)$ is the point mass function at $a$. 
	
	The PN model in \eqref{eq:y} is motivated by \cite{mehrpouyan2012joint}. In particular, the phase evolution of free-running oscillators can be modeled as a Wiener process (Brownian motion). Therefore, $\theta_i(t)$, $i=1,\ldots,K$, and $\phi_{m}(t)$, with $m = 1,\ldots,M$, are expressed as
	\begin{align} \label{eq:Wiener}
		\theta_{i}(t) &= \theta_i(t-1) + \Delta\theta_i(t), \nonumber\\
		\phi_m(t) &= \phi_m(t-1) + \Delta\phi_m(t),
	\end{align}
	where the phase innovations for the $i$-th user $\Delta\theta_i(t)$ and $m$-th receive antennas $\Delta\phi_m(t)$ are assumed to be white Gaussian processes, i.e., $\Delta\theta_i(t) \sim \mathcal{N}\!\big(0,\sigma^2_{\mr{t},i}\big)$ and $\phi_m(t) \sim \mathcal{N}\!\big(0,\sigma^2_{\mr{r},m}\big)$, respectively. 
	Throughout this paper, we assume that $\sigma^2_{\mr{t},i}$ and $\sigma^2_{\mr{r},m}$ are known at the receiver, given that they are dependent on the oscillator properties. With small $\sigma^2_{\mr{t},i}$ and $\sigma^2_{\mr{r},m}$,  the phase innovations $\Delta\theta_i(t)$ and $\Delta\phi_m(t)$ can be approximated by von Mises distributions as
\begin{align}
    \Delta\theta_i(t) \sim \mc{VM}(0,\varepsilon_{\mr{t},i}), \quad
    \Delta\phi_m(t) \sim \mathcal{VM}\big(0,\varepsilon_{\mr{r},m}),
\end{align}
where $\varepsilon_{\mr{t},i} = 1/\sigma^2_{\mr{t},i}$ and $\varepsilon_{\mr{r},m} = 1/\sigma^2_{\mr{r},m}$. Thus, 
\begin{align} 
p\big(\theta_i(t)\!\mid\!\theta_i(t-1)\big) &\approx \mc{VM}\big(\theta_i(t);\theta_i(t-1),\varepsilon_{\mr{t},i}\big),\\
p\big(\phi_m(t)\!\mid\!\phi_m(t-1)\big) &\approx \mc{VM}\big(\phi_m(t); \phi_m(t-1),\varepsilon_{\mr{r},m}\big).
\end{align} 
In the data transmission period, each transmitted symbol $s_{i}(t)$ is perturbed by a PN component $\theta_i(t)$, such that the \textit{phase-noise-corrupted transmit signal} from user-$i$ is given by
\begin{align} \label{eq:x_i}
	x_i(t) = s_{i}(t) \e^{\ji\theta_i(t)}.
\end{align}
Let $\mb{x}(t) = \bs{\Theta}(t)\mb{s}(t)$ be the PN-corrupted transmit vector; the received signal vector at the BS can be expressed as
\begin{align}\label{eq:system_model}
    {\mb{y}}(t) = \bs{\Phi}(t)\mb{H} {\mb{x}}(t) + {\mb{n}}(t).
\end{align}
\subsection{Relevant Results on von Mises Distribution} \label{sec:preliminaries}
    \begin{itemize}
    \item For a random variable (RV) ${x} \sim \mc{VM}(\mu,\kappa)$, the mean of  $\e^{\ji x}$ (i.e., the first complex trigonometric moment or mean resultant vector) and its variance are given by
    \begin{align} \label{eq:cir}
        \E\big[\e^{\ji {x}}\big] = \frac{I_{1} (\kappa)}{I_0(\kappa)} \e^{\ji \mu}, \quad \mr{Var}[\e^{\ji x}] = 1 - \left(\frac{I_{1} (\kappa)}{I_0(\kappa)}\right)^2,
    \end{align}
    where $I_i(\kappa)$ is the modified Bessel function of the first kind of order $i$. For notational simplicity, we define the ratio $I_1(\kappa)/I_0(\kappa)$ by a function $J(\kappa)$.
    \item Let $x \sim \mathcal{VM}(\mu_x, \kappa_x)$ and $n \sim \mathcal{VM}(0, \kappa_n)$ be independent,  and define $y = x + n \;(\mr{mod }\; 2\pi)$. While $y$ is \emph{not} von Mises in general, its density can be approximated by $\mc{VM}(\mu_y,\kappa_y)$. Matching the mean direction and the first trigonometric moment to a $\mc{VM}(\mu_y,\kappa_y)$ yields $\mu_y = \mu_x$ and $J(\kappa_y) = J(\kappa_x)J(\kappa_n)$, and thus $\kappa_y = J^{-1}\big(J(\kappa_x)J(\kappa_n)\big)$. When both $\kappa_x \gg 1$ and $\kappa_n \gg 1$, $\kappa_y \approx \frac{\kappa_x \kappa_n}{\kappa_x + \kappa_n}$. 		
\end{itemize} 
	
%
%
\begin{figure*}
  	\centering
 	\includegraphics[width=0.8\linewidth]{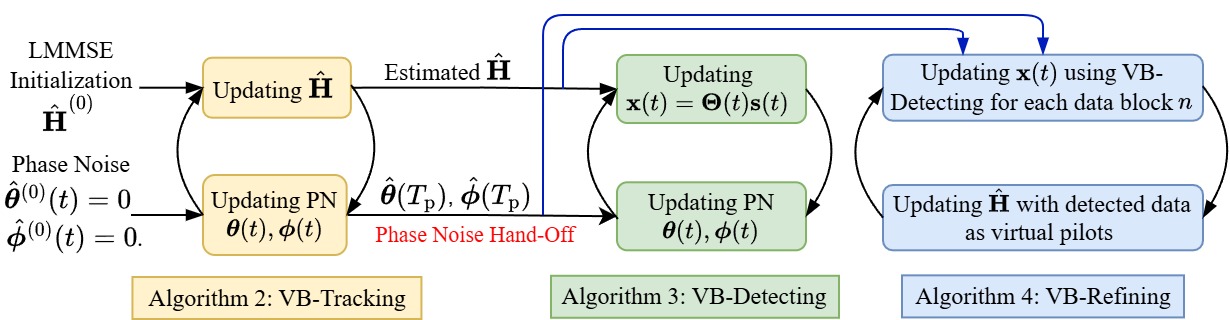}
 	\caption{Block diagram of the proposed channel estimation, data detection, phase noise tracking, and channel refinement framework.}
 	\label{fig:Block_diagram}
\end{figure*}	
	
\section{Channel Estimation and Phase Tracking}
\label{sec:VB:CE}
The proposed VB framework is sketched in Fig.~\ref{fig:Block_diagram}. VB-Tracking (Algorithm~\ref{algo:VB-Tracking}) jointly estimates the channel and PN trajectories during the pilot phase, starting from the LMMSE estimate, and hands the channel and terminal PN posteriors to VB-Detecting (Algorithm~\ref{algo:VB-Detecting}). VB-Refining (Algorithm~\ref{algo:VB-Refining}) runs VB-Detecting block by block and uses reliably detected symbols as virtual pilots to update the channel between blocks.

\subsection{Channel Initialization}
\label{sec:two_stage}
To initialize the VB joint estimator in Section~\ref{sec:vb_channel}, we first obtain a rough channel estimate via linear MMSE (LMMSE) while ignoring PN. This means we initialize the phase noise trajectories uninformatively, i.e., $\hat{\phi}_m(t)=0$ and $\hat{\theta}_i(t) = 0$, $\forall t = 1,\ldots,T_\p$, with concentration parameters set to the raw innovation values $\kappa_{\mr{t},i}(t) = \varepsilon_{\rm{t},i}$ and $\kappa_{\mr{r},m}(t) = \varepsilon_{\rm{r},m}$, and let Algorithm~\ref{algo:VB-Tracking} refine the channel and phase estimates jointly from there.
\subsubsection{LMMSE Channel Estimation}
While the PN is present but ignored by setting $\bm{\Phi}(t)=\mathbf{I}_M$ and $\bm{\Theta}(t)=\mathbf{I}_K$ for all $t$, which is the same initialization strategy adopted in \cite{wang2025iterative}, the received signal during the pilot simplifies to
\begin{equation}
	\mathbf{y}(t) = \mathbf{H}\mathbf{s}(t) + \mathbf{n}(t), 
	\quad t = 1,\ldots,T_\p.
\end{equation}	
Stacking all $T_\p$ observations column-wise gives
\begin{equation}
	\mathbf{Y}_\p = \mathbf{H}\mathbf{S}_\p + \mathbf{N}_\p,
\end{equation}
where $\mathbf{Y}_\p = [\mathbf{y}(1),\ldots,\mathbf{y}(T_\p)] \in \mathbb{C}^{M\times T_\p}$,
$\mathbf{S}_\p = [\mathbf{s}(1),\ldots,\mathbf{s}(T_\p)] \in \mathbb{C}^{K\times T_\p}$, and $\mathbf{N}_\p \in \mathbb{C}^{M\times T_\p}$ has i.i.d.\ $\mathcal{CN}(0,N_0)$ entries.  
The matched-filtered  observation for user $i$ under orthogonal pilots 	$\mathbf{S}_\p\mathbf{S}_\p^H=\rho_\p\mathbf{I}_K$ is
	\begin{equation}
		\tilde{\mathbf{y}}_i
		= \mathbf{Y}_\p\mathbf{s}_i^*
		= \rho_\p\mathbf{h}_i + \tilde{\mathbf{n}}_i,
	\end{equation}
	where $\tilde{\mathbf{n}}_i=\mathbf{N}_\p\mathbf{s}_i^*\sim 	\mathcal{CN}(\mathbf{0},\rho_\p N_0\mathbf{I}_M)$ and $\rho_\p = \|\mathbf{s}_i\|^2$. The $m$-th element satisfies
	\begin{equation}
		[\tilde{\mathbf{y}}_i]_m  = h_{mi}\,\rho_\p\cdot
		\underbrace{\frac{1}{T_\p}\sum_{t=1}^{T_\p}
			\e^{\ji(\phi_m(t)+\theta_i(t))}}_{\text{phase noise averaging factor}}
		+ [\mathbf{N}_\p\mathbf{s}_i^*]_m.
	\end{equation}
	Since $\phi_m(t)$ and $\theta_i(t)$ are zero-mean Wiener processes with small	incremental variance, the averaging factor has magnitude close to unity, so the
	LMMSE estimate is a mildly attenuated and phase-rotated version of the true channel. Given the prior $\mathbf{h}_i\sim\mathcal{CN}(\mathbf{0},\mathbf{C}_i)$,
	the LMMSE estimator and its error covariance are
	\begin{align}
		\hat{\mb{h}}_i^{(0)}
		&= \gamma_\p\rho_\p\,\mathbf{C}_i
		\!\left(\gamma_\p\rho_\p\,\mathbf{C}_i+\mathbf{I}_M\right)^{-1}
		\frac{\tilde{\mathbf{y}}_i}{\rho_\p},
		\label{eq:lmmse_mean} \\
		\bm{\Sigma}_i^{\text{LMMSE}}
		&= \left(\mathbf{C}_i^{-1}+\gamma_\p\rho_\p\,\mathbf{I}_M\right)^{-1},
		\label{eq:lmmse_cov}
	\end{align}
	where $\gamma_\p=1/N_0$ is the noise precision.
\subsubsection{Identifiability and Reference Phase} \label{subsubsec:Identifiability}
At $t=1$, the received signal is
	$\mathbf{y}(1)=\boldsymbol{\Phi}(1)\mathbf{H}\boldsymbol{\Theta}(1)
	\mathbf{s}(1)+\mathbf{n}(1)$.
	The LMMSE estimator observes only the product
	\begin{equation}
		\tilde{h}_{mi} \triangleq \e^{\ji\phi_m(1)}h_{mi} 
		\e^{\ji\theta_i(1)},
		\label{eq:htilde}
	\end{equation}
	so the combined phase $\phi_m(1)+\theta_i(1)$ is absorbed into each
	coefficient of the effective channel $\tilde{\mathbf{H}} = \boldsymbol{\Phi}(1)\mathbf{H}\boldsymbol{\Theta}(1)$.
	The $M+K$ individual initial phases cannot be separated from the $MK$ channel coefficients; only a single global phase is unidentifiable from $\mathbf{y}(1)$ alone. We therefore set $\phi_1(1)=0$ as the reference and absorb all remaining initial phases $\{\phi_m(1)\}_{m=2}^{M}$ and $\{\theta_i(1)\}_{i=1}^{K}$ into $\tilde{\mathbf{H}}$. This convention is valid throughout the paper provided: (i) the channel prior $p(\mb{h}_i; \mb{C}_i)$ is invariant under left and right unitary diagonal scaling, which holds under i.i.d. Rayleigh fading; (ii) the estimated $\tilde{\mathbf{H}}$ is used consistently in both pilot and data phases; and (iii) the Wiener process for $t \geq 2$ is initialized at $\theta_i(1) = \phi_m(1) = 0$, $\forall i,m$ so that all subsequent phase drift is measured relative to the absorbed initial value.
    For notational convenience we write $\hat{\mb{H}}^{(0)}$ in place of $\tilde{\mb{H}}$ hereafter and $\mb{H}$ in place of $\tilde{\mb{H}}$ from Section \ref{sec:vb_channel} onward.
\subsection{Variational Bayes Channel Estimation and Phase Tracking} \label{sec:vb_channel}
We aim to estimate the channel and phase noises during the pilot transmission. Denote $\gamma_\p$ as the precision of the noise in the pilot transmission phase, $\{\mb{C}_i\}$ as the set of covariance matrices $\mb{C}_1,\ldots,\mb{C}_K$, $\bs{\theta}(t) = \{\theta_i(t)\}_{i=1}^K$, and $\bs{\phi}(t) = \{\phi_m(t)\}_{m=1}^M$. Following the identifiability convention established in Subsection \ref{subsubsec:Identifiability}, we will use $\mb{H}$ in place of $\tilde{\mb{H}}$ with $\bs{\phi}(1) = \mb{0}$ and $\bs{\theta}(1) = \mb{0}$.
The conditional is given by
	\begin{align} \label{eq:con}
		&p(\mb{Y}_\p,\mb{H},\{\bs{\phi}(t)\},\{\bs{\theta}(t)\},\gamma_\p) \nonumber\\ 
		&=p(\mb{Y}_\p\!\mid\!\mb{H},\{\bs{\phi}(t)\},\{\bs{\theta}(t)\},\gamma_\p;\mb{S}_\p)p(\mb{H}|\{\mb{C}_i\})\nonumber\\
		&= \Bigg[\prod_{t=1}^{T_\p} p(\mb{y}(t)\!\mid\!\mb{H},\bs{\phi}(t),\bs{\theta}(t),\gamma_\p;\mb{s}(t))\Bigg]p(\gamma_\p)\Bigg[\prod_{i=1}^{K}p(\mb{h}_i;\mb{C}_i)\Bigg] \nonumber\\
		&\times \Bigg[\prod_{t=2}^{T_\p} p(\bs{\phi}(t)\!\mid\!\bs{\phi}(t-1))\Bigg] \Bigg[\prod_{t=2}^{T_\p} p(\bs{\theta}(t)\!\mid\!\bs{\theta}(t-1))\Bigg] ,
	\end{align}
	where 
	\begin{align} 
    &p\big(\mb{y}(t)\!\mid\!\mb{H},\bs{\phi}(t),\bs{\theta}(t),\gamma_\p;\mb{s}(t)\big)\nonumber \\
		&=\mc{CN}\big(\bs{\Phi}^H(t)\mb{y}(t);\mb{H}\bs{\Theta}(t)\mb{s}(t) ,\gamma_\p^{-1}\mb{I}_M\big).
    \end{align}
   
	We approximate the posterior with a mean-field variational distribution:
	\begin{align} \label{eq:mean-field:channel}
		&p(\mb{H},\bs{\phi}(t),\bs{\theta}(t),\gamma_\p \!\mid\! \mb{Y}_\p;\mb{S}_\p) \nonumber \\
        &\approx q(\mb{H},\bs{\phi}(t),\bs{\theta}(t),\gamma_\p)\nonumber\\
		&= \!\Bigg[\!\prod_{i=1}^K \!q(\mb{h}_i) \!\Bigg]\! q(\gamma_\p)  \Bigg[\prod_{t=2}^{T_\p}\prod_{i=1}^K \! q(\theta_i(t)) \!\Bigg] \! \Bigg[\prod_{t=2}^{T_\p}\prod_{m=1}^M \!q(\phi_m(t)) \!\Bigg].
	\end{align}
	
	For a \emph{current} estimate of the variational distributions $q(\mb{h}_i)$, $q(\bs{\phi}(t))$, $q(\bs{\theta}(t))$, and $q(\gamma_\p)$, we shall obtain the corresponding variational means $\hat{\mb{h}}_i$, $\hat{\bs{\phi}}(t)$, $\hat{\bs{\theta}}(t)$, and $\hat{\gamma}_\p$, respectively. For notational simplicity and computational efficiency, we define residual terms $\mb{r}(t) =  \hat{\bs{\Phi}}(t)^H\mb{y}(t)  - \hat{\mb{H}}\hat{\bs{\Theta}}(t){\mb{s}}(t)$ using these \emph{current} corresponding variational means.

\emph{1.) Update $\phi_m(t)$:} For $t=3,\ldots,T_\p-1$, by taking the expectation of the conditional \eqref{eq:con} w.r.t. to all latent variables except $\phi_m(t)$,  we obtain the variational distribution $q(\phi_m(t))$ as in \eqref{eq:phi:pilot}, 
\begin{figure*}
    \begin{align} \label{eq:phi:pilot}
        q(\phi_m(t)) &\propto \exp\Big\{\blr{\ln p\big(\phi_m(t)\!\mid\!\phi_m(t-1)\big) + \ln p\big(\phi_{m}(t+1)\!\mid\!\phi_m(t)\big) + \ln p\big(\mb{y}(t)\!\mid\!\mb{H},\bs{\phi}(t),\bs{\theta}(t), \gamma_\p;\mb{s}(t)\big)}_{-\phi_m(t)}\Big\}\nonumber\\
        &\propto \exp\Big\{\blr{ \Re\big\{\varepsilon_{\mr{r},m}^*\e^{\ji(\phi_m(t)-\phi_m(t-1))}\big\} + \Re\big\{\varepsilon_{\mr{r},m}^*\e^{\ji(\phi_m(t+1)-\phi_m(t))}\big\} -{\gamma}_\p\big|\e^{-\ji \phi_m(t)}y_m(t) - \mb{H}_{m,:}\bs{\Theta}(t)\mb{s}(t)\big|^2}_{-\phi_m(t)}\Big\}\nonumber\\
        &\propto \exp\Big\{ \Re\big\{\varepsilon_{\mr{r},m}^*\blr{ \e^{-\ji\phi_m(t-1)} }  \e^{\ji \phi_m(t)} \big\}  
        + \Re\big\{\varepsilon_{\mr{r},m}^*\blr{\e^{\ji\phi_m(t+1)}}\e^{-\ji \phi_m(t)}\big\} + 2\hat{\gamma}_\p\Re\{\e^{\ji \phi_m(t)}y_m^*(t)\mb{H}_{m,:}\bs{\Theta}(t) \mb{s}(t)\}\Big\} \nonumber\\
        &\propto \exp\left(\Re\big\{\eta_{\mr{r},m}^*\e^{\ji \phi_m(t)}\big\}\right),			
    \end{align}
    \hrule
\end{figure*}  
where $\nu_{\mr{r},m}(t) = 2\hat{\gamma}_\p y_m(t)(\mb{H}_{m,:}\bs{\Theta}(t) \mb{s}(t))^* $  	and $\eta_{\mr{r},m}(t) = \varepsilon_{\mr{r},m} \big(\blr{\e^{\ji\phi_m(t-1)}}  + \blr{\e^{\ji\hat{\phi}_m(t+1)}}\big) + \nu_{\mr{r},m}(t)$. Thus, $q\big(\phi_m(t)\big)$ is a von Mises density $\mc{VM}\big(\phi_m(t);\eta_{\mr{r},m}(t)\big)$. 
	
	We then can obtain the mean direction $\hat{\phi}_{m}(t) = \angle \eta_{\mr{r},m}(t)$ and the  resultant vector $\blr{\e^{\ji \phi_{m}(t)}} = J\big(|\eta_{\mr{r},m}(t)|\big)\e^{\ji \hat{\phi}_{m}(t)}$. Similarly,  $q\big(\phi_{m}(t-1)\big)$ and $q\big(\phi_{m}(t+1)\big)$ are von Mises densities, enabling the computations of $\blr{\e^{\ji \phi_{m}(t-1)}}$ and $\blr{\e^{\ji \phi_{m}(t+1)}}$. Note that $\nu_{\mr{r},m}(t)$ can be computed efficiently using the \emph{current} variational density of $\phi_m(t)$ and the residual term $r_m(t)$ as
		\begin{align} 
	\nu_{\mr{r},m}(t) &= 2\hat{\gamma}_\p y_m(t)(\mb{H}_{m,:}\bs{\Theta}(t) \mb{s}(t))^* \nonumber \\&= 2\hat{\gamma}_\p y_m(t)\big(\blr{\e^{-\ji\phi_m(t)}}y_m(t) - r_m(t)\big)^* \nonumber \\
	&=  2\hat{\gamma}_\p \big(|y_m(t)|^2 J\big(|\eta_{\mr{r},m}(t)|\big) \e^{\ji\hat{\phi}_m(t)} -y_m(t) r_m^*(t)\big).
\end{align}
	The definition of $\nu_{\mr{r},m}(t)$ is also valid for $t=2$ and $t=T_\p$. 

	For $t=2$, we use the prior $p\big(\phi_m(2)\big) = \mc{VM}\big(\phi_m(2);\varepsilon_{\mr{r},m}\big)$ in lieu of the conditional $p\big(\phi_m(2)\!\mid\! \phi_m(1)\big)$. The variational distribution $q\big(\phi_m(2)\big)$ is thus von Mises with parameter $\eta_{\mr{r},m}(2) = \varepsilon_{\mr{r},m}\big(1  +  J\big(|\eta_{\mr{r},m}(3)|\big)\e^{\ji\hat{\phi}_m(3)}\big) + \nu_{\mr{r},m}(2)$.
	
	For $t=T_\p$, since the conditional probability $p\big(\phi_m(T_\p + 1) \!\mid\! \phi_m(T_\p)\big)$  is not available, the variational distribution $q\big(\phi_{m}(T_\p)\big)$ is thus von Mises with parameter $\eta_{\mr{r},m}(T_\p) = \varepsilon_{\mr{r},m}J\big( |\eta_{\mr{r},m}(T_\p-1)|\big) \e^{\ji\hat{\phi}_m(T_\p-1)} +  \nu_{\mr{r},m}(T_\p)$.
	
	\emph{2.) Update $\theta_i(t)$:}  For $t=3,\ldots,T_\p-1$, by taking the expectation of the conditional \eqref{eq:con} w.r.t. to all latent variables except $\theta_i(t)$, we obtain the variational distribution $q(\theta_i(t))$ as von Mises in \eqref{eq:theta}, 
	\begin{figure*} [!thp]
    \vspace{-0.4cm}
		\begin{align} \label{eq:theta}
			q(\theta_i(t)) &\propto \exp\Big\{\blr{\ln p\big(\theta_i(t)\!\mid\!\theta_i(t-1)\big) + \ln p\big(\theta_i(t+1)\!\mid\!\theta_i(t)\big) + \ln p\big(\mb{y}(t)\!\mid\!\mb{H},\bs{\phi}(t),\bs{\theta}(t), \gamma_\p;\mb{s}(t)\big)}_{-\theta_i(t)}\Big\}\nonumber\\
			&\propto \exp\Big\{ \blr{\Re\big\{\varepsilon_{\mr{t},i}^*\e^{\ji(\theta_i(t)-\theta_i(t-1))}\big\} 
			+ \Re\big\{\varepsilon_{\mr{t},i}^*\e^{\ji(\theta_i(t+1)-\theta_i(t))}\big\} 
			- \gamma_\p \big\|\bs{\Phi}^H(t)\mb{y}(t) - \mb{H}\mb{\Theta}(t)\mb{s}(t) \big\|^2}_{-\theta_i(t)} \Big\} \nonumber \\
			&\propto \exp\bigg\{ \Re\big\{\varepsilon_{\mr{t},i}^* \blr{\e^{-\ji\theta_i(t-1)}} \e^{\ji\theta_i(t)}\big\} 
			+ \Re\big\{\varepsilon_{\mr{t},i}^*\blr{\e^{\ji(\theta_i(t+1)}}\e^{-\ji \theta_i(t))}\big\} \nonumber \\
			&\quad\quad\quad  + 2\hat{\gamma}_\p\Re\bigg\{\bigg(\hat{\bs{\Phi}}^H(t)\mb{y}(t) - \sum_{j\neq i}^K \hat{\mb{h}}_j\blr{\e^{\ji \theta_i(t))}}s_j(t)\!\bigg)^H \!\hat{\mb{h}}_i {s}_i(t)\e^{\ji \theta_i(t)}\bigg\}\bigg\} \nonumber  \\
			&\propto \exp\left(\Re\big\{\eta_{\mr{t},i}^*\e^{\ji \theta_i(t)}\big\}\right),
		\end{align}
		\hrule
	\end{figure*} 
	where \begin{align}
	    \nu_{\mr{t},i}(t) = 2\hat{\gamma}_\p\hat{\mb{h}}_i^H\left(\hat{\bs{\Phi}}(t)^H\mb{y}(t) - \sum_{j\neq i}^K \hat{\mb{h}}_j\blr{\e^{\ji \theta_j(t)}}{s}_j(t)\right) s_i^*(t),
	\end{align} 
	and $\eta_{\mr{r},i}(t) = \varepsilon_{\mr{t},i}\big(\blr{\e^{\ji\hat{\theta}_i(t-1)}} + \blr{ \e^{\ji\hat{\theta}_i(t+1)}} \big)+ \nu_{\mr{t},i}(t)$. Thus, $q\big(\theta_i(t)\big)$ is a von Mises density $\mc{VM}\big(\theta_i(t);\eta_{\mr{t},i}(t)\big)$. 
	
	We then can obtain the mean direction $\hat{\theta}_{i}(t) = \angle \eta_{\mr{t},i}(t)$ and the resultant vector $\blr{\e^{\ji \theta_{i}(t)}} = J\big(|\eta_{\mr{t},i}(t)|\big)\e^{\ji \hat{\theta}_i(t)}$. Similarly,  $q\big(\theta_i(t-1)\big)$ and $q\big(\theta_{i}(t+1)\big)$ are von Mises densities, enabling the computations of $\blr{\e^{\ji \theta_{i}(t-1)}}$ and $\blr{\e^{\ji \theta_{i}(t+1)}}$. Note that $\nu_{\mr{t},i}(t)$ can be computed efficiently using the \emph{current} variational density of $\theta_i(t)$ and the residual term $\mb{r}(t)$ as	
	\begin{align}
		\nu_{\mr{t},i}(t) &= 2\hat{\gamma}_\p\hat{\mb{h}}_i^H\left(\hat{\bs{\Phi}}(t)^H\mb{y}(t) - \sum_{j\neq i}^K \hat{\mb{h}}_j\blr{\e^{\ji \theta_j(t)}}{s}_j(t)\right) s_i^*(t)\nonumber\\
		&= 2\hat{\gamma}_\p \hat{\mb{h}}_i^H\left(\mb{r}(t) +  \hat{\mb{h}}_i\blr{\e^{\ji \theta_i(t)}}s_i(t)\right)s_i^*(t) \nonumber\\
		&=  2\hat{\gamma}_\p \big(\|\hat{\mb{h}}_i\|^2 J\big(|\eta_{\mr{t},i}(t)|\big) \e^{\ji \hat{\theta}_{i}(t)}|s_i(t)|^2 + \hat{\mb{h}}_i^H \mb{r}(t)s_i^*(t) \big).
	\end{align}
	The definition of $\nu_{\mr{t},i}(t)$ is also valid for $t=2$ and $t=T_\p$. 
	
	For $t=2$, we use the prior $p\big(\theta_i(2)\big) = \mc{VM}\big(\theta_i(2);\varepsilon_{\mr{t},i}\big)$  in lieu of the conditional $p\big(\theta_i(2)\!\mid\! \theta_i(1)\big)$. The variational distribution $q\big(\theta_i(2)\big)$ is also von Mises with parameter $\eta_{\mr{t},i}(t) = \varepsilon_{\mr{t},i}\big(1 +J\big(|\eta_{\mr{t},i}(3)|\big) \e^{\ji\hat{\theta}_i(3)} \big)+ \nu_{\mr{t},i}(2)$.
	
	For $t=T_\p$, since the conditional probability $p\big(\theta_i(T_\p + 1) \!\mid\! \theta_i(T_\p)\big)$  is not available, the variational distribution $q\big(\theta_i(T_\p)\big)$ is thus von Mises with parameter $\nu_{\mr{t},i}(T_\p) = \varepsilon_{\mr{t},i}J\big(|\eta_{\mr{t},i}(T_\p-1)|\big)  \e^{\ji\hat{\theta}_i(T_\p-1)} +  \eta_{\mr{t},i}(T_\p)$.

\emph{3.) Update $\mb{h}_i$:} Taking the expectation of the conditional \eqref{eq:con} w.r.t. all latent variables except $\mb{h}_i$, the variational distribution $q(\mb{h}_i)$ can be derived as
\begin{align}  \label{eq:h:0}
    q(\mb{h}_i) \propto \exp\bigg\{\bblr{ \sum_{t=1}^{T_\p}& \ln p\big(\mb{y}(t)\!\mid\!\mb{H},  \bs{\phi}(t),\bs{\theta}(t), \gamma_\p;\mb{s}(t) \big) \nonumber \\
        & + \ln p({\mb{h}_i};\mb{C}_i) }_{-\mb{h}_i}\bigg\},
\end{align}
which is further expanded in \eqref{eq:h:CE} in the page. %
\begin{figure*} 
    \vspace{-0.5cm}
\begin{align} 
	q(\mb{h}_i) &\propto \exp\Big\{\!\!-\Blr{\gamma_\p\sum_{t=1}^{T_\p}\|\bs{\Phi}^H(t)\mb{y}(t) - \mb{H}\bs{\Theta}(t)\mb{s}(t)\|^2 + \mb{h}_i^H\mb{C}_i^{-1}\mb{h}_i}\Big\}\nonumber\\
	&\propto \exp\Bigg\{\!\!- \hat{\gamma}_\p\sum_{t=1}^{T_\p} \Big(\|\mb{h}_i\|^2 |s_i(t)|^2 -2\Re\Big\{\mb{h}_i^H \blr{\e^{-\ji \theta_i(t)}} \Big(\hat{\bs{\Phi}}^H(t)\mb{y}(t) - \sum_{j\neq i}^K\lr{\mb{h}_j}\blr{\e^{\ji \theta_j(t)}}s_j(t)\Big)s^*_i(t) \Big\}\Big) - \mb{h}_i^H\mb{C}_i^{-1}\mb{h}_i\Big\}\nonumber\\
	&\propto \exp\Bigg\{\!\!-\!\mb{h}_i^H\Bigg[ \hat{\gamma}_\p \!\sum_{t=1}^{T_\p}\!  |s_i(t)|^2\mb{I}_M  {+} \mb{C}_i^{-1} \Bigg] \mb{h}_i {+} 2  \hat{\gamma}_\p \Re\bigg\{\mb{h}_i^H \sum_{t=1}^{T_\p}\blr{\e^{-\ji \theta_i(t)}} \Big(\!\hat{\bs{\Phi}}^H(t)\mb{y}(t) {-} \sum_{\ji\neq i}^K\hat{\mb{h}}_j \blr{\e^{\ji \theta_j(t)}}s_j(t)\Big)s_i^*(t) \bigg\} \Bigg\}, \label{eq:h:CE} 
			%
	\end{align}
	\hrule
\end{figure*} 
The variational variance and mean of $\mb{h}_i$ are given, respectively, as 
\begin{align}
\bs{\Sigma}_{\mb{h}_i} &= \Bigg[\hat{\gamma}_\p\sum_{t=1}^{T_\p}  |s_i(t)|^2 \mb{I}_M + \mb{C}_i^{-1} \Bigg]^{-1}, \label{eq:h:variance}\\
\lr{\mb{h}_i} &= \bs{\Sigma}_{\mb{h}_i}\hat{\gamma}_\p \sum_{t=1}^{T_\p}\Big[J\big(|\eta_{\mr{t},i}(t)|\big)\e^{-\ji \hat{\theta}_i(t)}\mb{r}(t) s_i^*(t) \nonumber\\
& \qquad\qquad + J\big(|\eta_{\mr{t},i}(t)|\big)^2 \hat{\mb{h}}_i |s_i(t)|^2 \Big], \label{eq:h:mean}
\end{align}  
where $$\blr{\e^{\pm \ji \theta_i(t)}} = J\big(|\eta_{\mr{t},i}(t)|\big) \e^{\pm \ji \hat{\theta}_i(t)}, \big|\blr{\e^{\ji \theta_i(t)}}\big|^2 = J\big(|\eta_{\mr{t},i}(t)|\big)^2.$$

\emph{4.) Update $\gamma_\p$:} 
The variational distribution $q(\gamma_\p)$ can be obtained as 
\begin{align}  \label{eq:q_nu_p}
q(\gamma_\p) &\propto \exp\Big\{\blr{ \ln p({\mb{Y}_{\p}}|\mb{H},\{\bs{\phi}(t)\},\{\bs{\theta}(t)\},\gamma_\p; \mb{S}_{\p}) \nonumber\\
&\qquad\quad + \ln p(\gamma_\p) }\Big\} \nonumber\\
&\propto \exp\Big\{(a_\p + M T_\p - 1)\ln\gamma_\p - b_\p\gamma_\p  \nonumber\\
&\qquad\quad -\gamma_\p\!\sum_{t=1}^{T_\p}\!\blr{\|\bs{\Phi}^H(t)\mb{y}(t) - \mb{H} \bs{\Theta}(t)\mb{s}(t) \|^2}\Big\}.
\end{align}

Here, $q(\gamma_\p)$ is a Gamma distribution with mean
\begin{align} \label{eq:gamma_p}
    \lr{\gamma_\p} = \frac{a_\p + M T_\p}{b_\p + \sum_{t=1}^{T_\p}\blr{\|\bs{\Phi}^H(t)\mb{y}(t) - \mb{H} \bs{\Theta}(t)\mb{s}(t) \|^2 } },
\end{align}
where 
\begin{align} \label{eq:gamma_p:expand}
    &\blr{\| \bs{\Phi}(t)^H\mb{y}(t) - \mb{H} \bs{\Theta}(t)\mb{s}(t) \|^2 }\nonumber\\
      &= \|\mb{r}(t)\|^2 + \sum_{m=1}^M|y_m(t)|^2\big[1-J(\eta_{\mr{r},m}(t))^2\big]  \nonumber\\
    & + \sum_{i=1}^K |s_i(t)|^2 \Big[ \tr\{\mb{\Sigma}_{\mb{h}_i} \} + \|\lr{\mb{h}_i}\|^2 (1-J(\eta_{\mr{t},i}(t))^2)  \Big],
\end{align}
and $\mb{r}(t) = \lr{\bs{\Phi}(t)}^H\mb{y}(t) \!-\! \lr{\mb{H}}\lr{\bs{\Theta}(t)}  \mb{s}(t)$, which is also a column of the residual matrix $\mb{R}_\p$, i.e., $\mb{r}(t) = [\mb{R}_{\p}]_{:,t}$.

\begin{algorithm}[t]
\small
\caption{-- Joint Channel Estimation and Phase Noise Tracking}
\label{algo:VB-Tracking}
\begin{algorithmic}[1]
\State \textbf{Input:} $(\mb{Y}_\p,\mb{S}_\p)$, $\{\mb{C}_i\}$, $\{\varepsilon_{\mr{t},i}\}$, $\{\varepsilon_{\mr{r},m}\}$, $(a_\p,b_\p)$.
\State \textbf{Output:} $\hat{\mb{H}}$, $\{\hat{\phi}_m(t),\kappa_{\mr{r},m}(t)\}$, $\{\hat{\theta}_i(t),\kappa_{\mr{t},i}(t)\}$, $\hat{\gamma}_\p$
\State Initialize $\hat{\mb{h}}_i \gets \hat{\mb{h}}_i^{(0)}$ and $\bs{\Sigma}_{\mb{h}_i} \gets \bs{\Sigma}_i^{\text{LMMSE}}$, $\forall i$ based on \eqref{eq:lmmse_mean} and \eqref{eq:lmmse_cov}, respectively.
\State Initialize $\hat{\theta}_i(t) = 0$, $\hat{\phi}_m(t)=0$, $i=1,\ldots,K, \, m = 1,\ldots,M,\,t=1,\ldots,T_\p.$
\State Initialize $\eta_{\mr{r},m}(t) \gets \kappa_{\mr{r},m}(t), \eta_{\mr{t},i}(t) \gets \kappa_{\mr{t},i}(t)$,  ${[\mb{R}_\p]_{:,t}
        \gets \hat{\bs{\Phi}}(t)^H\mb{y}(t)
                   - \hat{\mb{H}}^{(0)}\hat{\bs{\Theta}}(t)\mb{s}(t)}$, $\forall\,m,\,t$
\For{$\ell = 1,2,\ldots$}
\State Update $\hat{\gamma}^\ell_\p$ via~\eqref{eq:gamma_p} and \eqref{eq:gamma_p:expand} 
    \For{$i = 1,\ldots,K$}
        \State $\mb{g}_{i}^\ell \gets \big[J(|\eta^\ell_{\mr{t},i}(1)|)\e^{\ji \hat{\theta}_{i}^\ell(1)},\ldots, J(|\eta^\ell_{\mr{t},i}(T_\p)|)\e^{\ji \hat{\theta}_{i}^\ell(T_\p)}\big]^T$
        \State $\mb{z}_i^\ell \gets \mb{g}_{i}^\ell \odot [\mb{S}_{\p}]_{i,:}^T$
        \State $\bs{\Sigma}_{\mb{h}_i}^{\ell+1} \gets \left[\hat{\gamma}_\p^\ell \sum_{t=1}^{T_\p}|s_i(t)|^2 \mb{I}_M + \mb{C}_i^{-1}\right]^{-1}$ 
		\State $\mb{h}_i^{\ell+1} \gets \bs{\Sigma}_{\mb{h}_i}^{\ell+1}\hat{\gamma}^\ell_\p [\mb{R}_\p (\mb{z}^\ell_i)^* + \|\mb{z}^\ell_i\|^2\hat{\mb{h}}_i^\ell]$
        \State $\mb{R}_\p \gets \mb{R}_\p + (\hat{\mb{h}}_i^\ell-\hat{\mb{h}}_i^{\ell+1}) (\mb{z}_i^\ell)^T$
    \EndFor
    \For{$t = 2,\ldots,T_\p$} \Comment{Update $q(\theta_i(t))$}
        \For{$i = 1,\ldots,K$}
            \State $\nu^\ell_{\mr{t},i}(t) \gets 2\hat{\gamma}^\ell_\p \big[\|\hat{\mb{h}}^{\ell+1}_i\|^2 J(|\eta_{\mr{t},i}(t)|)\e^{\ji\hat{\theta}_i(t)} |s_i(t)|^2 $\Statex \qquad \qquad\qquad\qquad\qquad $+ (\hat{\mb{h}}^{\ell+1}_i)^H\mb{r}(t) s^*_i(t)\big]$ 
            \State $\eta^\ell_{\mr{t},i}(t) \gets \varepsilon_{\mr{t},i}\big[J(|\eta^\ell_{\mr{t},i}(t-1)|)\e^{\ji\hat{\theta}^\ell_i(t-1)} $\Statex \qquad \qquad\qquad\qquad $ + J(|\eta^\ell_{\mr{t},i}(t+1)|)\e^{\ji\hat{\theta}^\ell_i(t+1)}\big] + \nu^\ell_{\mr{t},i}(t)$ 
            \State $\hat{\theta}_i^{\ell+1}(t) \gets \arg (\eta^\ell_{\mr{t},i}(t))$
            \State  $d_{\mr{t},i}^{\ell+1}(t) \gets J(|\eta^\ell_{\mr{t},i}(t)|)$
            \State $[\mb{R}_\p]_{:,t} \gets [\mb{R}_\p]_{:,t} + \hat{\mb{h}}_i^{\ell+1} {s}_i(t)$
            \Statex \qquad\qquad\quad $\times \big( d_{\mr{t},i}^{\ell}(t)\e^{\ji \hat{\theta}_{i}^{\ell}(t)}  - d_{\mr{t},i}^{\ell+1}(t)\e^{\ji \hat{\theta}_{i}^{\ell+1}(t)}\big)$
        \EndFor
        \State $\bs{\nu}_{\mr{r}}^\ell(t) \gets 2\hat{\gamma}^\ell_\p \big[\mb{y}(t)\odot \mb{y}^*(t) J(|\bs{\eta}_{\mr{r}}(t)|)\e^{\ji\hat{\bs{\phi}}^\ell(t)}$
            \Statex \qquad \qquad\qquad\qquad \qquad  $ - \mb{y}(t) \odot\mb{r}^*(t)\big]$  
            \State $\bs{\eta}^{\ell}_{\mr{r}}(t) \gets \bs{\varepsilon}^\ell_{\mr{r}}\big[J(|\bs{\eta}^{\ell}_{\mr{r}}(t-1)|)\e^{\ji\hat{\bs{\phi}}^\ell(t-1)}$
            \Statex \qquad \qquad\qquad\quad\quad  $ + J(|\bs{\eta}^{\ell}_{\mr{r}}(t+1)|)\e^{\ji\hat{\bs{\phi}}^\ell(t+1)}\big] + \bs{\nu}^\ell_{\mr{r}}(t)$ 
        \State $\hat{\bs{\phi}}^{\ell+1}(t) \gets \arg(\bs{\eta}^{\ell}_{\mr{r}}(t))$
        \State  $\mb{d}_{\mr{r}}^{\ell+1}(t) \gets J(|\bs{\eta}^{\ell}_{\mr{r}}(t)|)$
        \State $[\mb{R}_\p]_{:,t} \gets [\mb{R}_\p]_{:,t} -\mb{y}(t) \odot \big( \mb{d}_{\mr{r}}^{\ell}(t)\odot\e^{-\ji \hat{\bs{\phi}}^{\ell}(t)}$
        \Statex \qquad\qquad\qquad\qquad\qquad $  - \mb{d}_{\mr{r}}^{\ell+1}(t)\odot\e^{-\ji \hat{\bs{\phi}}^{\ell+1}(t)} \big)$\;
    \EndFor
\EndFor

\end{algorithmic}
\end{algorithm}

\subsection{Handoff to Data Detection}
At the end of the pilot block, Algorithm~2 delivers the terminal posterior estimates
$\{\hat{\phi}_m(T_\p),\kappa_{\mr{r},m}(T_\p)\}_{m=1}^{M}$ and
	$\{\hat{\theta}_i(T_\p),\kappa_{\mr{t},i}(T_\p)\}_{i=1}^{K}$.
Using the von Mises composition rule from Section~\ref{sec:preliminaries}, these are propagated one step forward to form the VM predictive priors entering data detection at $t=T_\p+1$:
	\begin{align}
		p\!\left(\phi_m(T_\p+1)\right)
		&= \mathcal{VM}\!\left(
		\hat{\phi}_m(T_\p),
		J^{-1}\!\left(
		J\!\left(\kappa_{\mr{r},m}(T_\p)\right) J(\varepsilon_{\mr{r},m})
		\right)
		\right),
		\label{eq:phi_pred_handoff} \\
		p\big(\theta_i(T_\p+1)\big)
		&= \mathcal{VM}\!\left(
		\hat{\theta}_i(T_\p),
		J^{-1}\left(
		J\big(\kappa_{\mr{t},i}(T_\p)\big) J\big(\varepsilon_{\mr{t},i}\big)
		\right)
		\right).
		\label{eq:theta_pred_handoff}
	\end{align}
These predictive priors initialize Algorithm 3, as detailed in the next Section.

\section{Variational Bayes for Joint Data Detection and Phase Tracking} \label{sec:DataDetection}
In this section, we develop a VB algorithm for data detection based on the estimated channel in Section \ref{sec:VB:CE}. The update process for the receiver's PN is the same as that in the pilot transmission. The only difference is that the transmitted symbol $s_i(t)$ is now unknown and must be jointly estimated with the PNs. With the fixed estimated channel $\hat{\mb{H}}$ from Section III, the received signal at slot $t = T_\p + 1,\ldots,T_\p +T_\da$ is
\begin{align}\label{eq:system_model}
    {\mb{y}}(t) = \bs{\Phi}(t)\hat{\mb{H}} \mb{x}(t) + {\mb{n}}(t),
\end{align}
where $\mb{x}(t) = \bs{\Theta}(t)\mb{s}(t)$ denotes the phase-noise-corrupted transmitted symbol vector. The posterior statistics of the received PN are unchanged, as derived in \eqref{eq:phi:pilot}, while those for the composite variable $x_{i}(t) = s_i(t)\e^{\ji \theta_i(t)}$ need to be derived. For ease of reference, we recall the following results from our previous work \cite{nguyen2026variational}, dropping the time index $t$ without loss of generality.
\begin{theorem}[Posteriors of $s$, $\theta$, and $x$ given $z$]
\label{thm:main}
Let ${s \in \mc{S}=\{s_1,\ldots,s_N\}\subset\mathbb{C}}$ be a discrete RV with prior probabilities $p_k = P(s = s_k)$.  Let $\theta \sim \mc{VM}(\varepsilon)$ be a phase noise RV with PDF $p(\theta;\varepsilon) \propto \exp\bigl(\Re\{\varepsilon^*\e^{\ji\theta}\}\bigr)$. 
The variables $s$ and $\theta$ are mutually independent. Let $z$ be a noisy observation of $x\triangleq s\,\e^{\ji\theta}$, i.e.,
\begin{equation}
	z = x + n = s\,e^{\ji\theta} + n,
	\quad n \sim \mathcal{CN}(0,\,\gamma^{-1}),
	\label{eq:model}
\end{equation}
where $n$ is independent of $s$ and $\theta$. For each $k \in \{1,\ldots,N\}$, define the \emph{posterior natural parameter}
\begin{equation}
\eta_k \triangleq \varepsilon + 2\gamma\,z\,s_k^*.	\label{eq:eta}
\end{equation}
The following results hold:

\noindent (I) \emph{Signal prior $p(x)$:} 
\begin{align}
	p(x) = \sum_{k=1}^N p_k\cdot \frac{\exp\big(\Re\{\varepsilon^* \e^{\ji (\angle x - \angle s_k)}\}\big)}{2\pi|s_k|I_0(|\varepsilon|)} \cdot \delta(|x| - |s_k|).
	\tag{I}
	\label{eq:x}
\end{align}
\noindent (II) \emph{Symbol posterior $p(s_k \!\mid\! z)$:}
\begin{equation}
\alpha_k \triangleq P(s = s_k \!\mid\! z) \propto p_k
		\cdot\exp\bigl(-\gamma\,|s_k|^2\bigr)
		\cdot I_0(|\eta_k|),
	\tag{II}
	\label{eq:ps}
\end{equation}
normalized so that $\sum_{k=1}^N \alpha_k = 1$.

\noindent (III) \emph{Phase posterior $p(\theta \!\mid\! z)$:}
\begin{equation}
	p(\theta \!\mid\! z) = \sum_{k=1}^{N} \alpha_k\cdot \mc{VM}(\theta;\eta_k), 
		\quad \theta\in[-\pi,\pi),
	\tag{III}
	\label{eq:ptheta}
\end{equation}
a finite mixture of von Mises distributions with a conditional mean 
\begin{equation}
	\mathbb{E}[\theta\!\mid\! z] = \scalebox{1.5}{$\angle$} \left(\sum_{k=1}^N \alpha_k \cdot \frac{I_1(|\eta_k|)}{I_0(|\eta_k|)} \cdot \e^{\ji\angle \eta_k}\right).
\end{equation}

\noindent(IV) \emph{Signal posterior $p(x \!\mid\! z)$:}
\begin{equation}
{p(x \!\mid\! z)
		= \sum_{k=1}^{N} \alpha_k \cdot
		\frac{\exp\bigl(\Re\bigl\{\eta_k^*\e^{\ji(\angle x - \angle s_k)}\bigr\}\bigr)}
		{2\pi\,|s_k|\,I_0(|\eta_k|)}
		\cdot \delta(|x| - |s_k|),}
	\tag{IV}
	\label{eq:px}
\end{equation}
a finite mixture of von Mises distributions on concentric circles of radii $\{|s_k|\}$.
\end{theorem}

\begin{corollary}[Posterior Mean and Variance of $x$]
	\label{cor:meanvar}
	Let $d_k \triangleq I_1(|\eta_k|)/I_0(|\eta_k|)$ be the \emph{mean resultant length} of the $k$-th component, and define the per-component posterior mean
	\begin{equation}
	 \mathbb{E}[x \!\mid\! z,\, s_k] = s_k\,d_k\,\e^{\ji\angle\eta_k}.
		\label{eq:xhat}
	\end{equation}
	
	\medskip\noindent (i) \emph{Posterior mean:}
	\begin{equation}
	\bar{x} =\mathbb{E}[x \!\mid\! z]
			= \sum_{k=1}^N \alpha_k\,s_k\,d_k\,\e^{\ji\angle\eta_k}.
		\label{eq:mean}
	\end{equation}
	Each symbol contributes its amplitude $|s_k|$, shrunk by the mean resultant length
	$d_k \in [0,1)$ and rotated to the posterior mean phase $\angle s_k + \angle\eta_k$.
	
	\medskip\noindent (ii) \emph{Posterior variance:}
	\begin{align}
		\mathrm{Var}[x \!\mid\! z]= \sum_{k=1}^N \alpha_k|s_k|^2 - |\bar{x}|^2.
		\label{eq:var}
	\end{align}
\end{corollary}
\begin{IEEEproof}
Please refer to \cite{nguyen2026variational} for the proof of Theorem \ref{thm:main} and Corollary \ref{cor:meanvar}.
\end{IEEEproof}

\begin{theorem} \label{thm:collapsedVM}
Let $p(\theta) = \sum_{k=1}^{N} \alpha_k \mc{VM}(\theta;\eta_k)$ for $\theta \in [-\pi, \pi)$ be a finite mixture of von Mises components with weights $\alpha_k \ge 0$, $\sum_{k=1}^N \alpha_k = 1$. The mixture can be approximated by a single von Mises distribution $\mc{VM}(\theta;\mu,\kappa)$ that minimizes the KL-divergence $D_{\mathrm{KL}}\big(p(\theta) \parallel \mc{VM}(\theta;\mu,\kappa)\big)$, where
\begin{align} 
    \mu &= \scalebox{1.5}{$\angle$} \bigg(\sum_{k=1}^N \alpha_k J(|\eta_k|) \e^{\ji\angle \eta_k}\bigg), \label{eq:mean:theta}\\
    \kappa &= J^{-1} \bigg(\bigg|\sum_{k=1}^N \alpha_k J(|\eta_k|) \e^{\ji\angle \eta_k}\bigg|\bigg),\label{eq:con:theta}
\end{align}
\end{theorem}

\begin{IEEEproof}
    See Appendix \ref{append:B}
\end{IEEEproof}

Joint data detection and phase tracking are performed sequentially as follows:
\subsubsection{Update $x_i(t)$} 
Since the channel is estimated and retained as a fixed posterior from Algorithm \ref{algo:VB-Tracking}, $q(\mb{h}_i) = \mc{CN}(\hat{\mb{h}}_i,\bs{\Sigma}_{\mb{h}_i})$,
the variational distribution $q(x_i(t))$ can be obtained as
\begin{align}  \label{eq:x_dt}
&q(x_i(t)) \nonumber\\
&\propto \exp\Big\{\blr{ \ln p({\mb{y}(t)}|\mb{H},\gamma_{\da}(t), \bs{\phi}(t), \mb{x}(t)) + \ln p(x_i(t)) }\Big\} \nonumber\\
&\propto p(x_i(t))\exp\Big\{- \hat{\gamma}_{\da}(t) \Big[\lr{ \|\mb{h}_i\|^2}|x_i(t)|^2 \nonumber\\
&\quad-2\Re\Big\{ \lr{\mb{h}_i^H} \Big( \hat{\bs{\Phi}}(t)^H\mb{y}(t) - \sum_{j\neq i}^K\lr{\mb{h}_j}\lr{x_{j}(t)} \Big) x_i(t)^*\Big\}\Big] \Big\} \nonumber\\
&\propto p(x_i(t))\exp\Big\{- \hat{\gamma}_{\da}(t) \lr{ \|\mb{h}_i\|^2}| z_{i}(t) - x_i(t)|^2 \Big\} \nonumber\\
&\propto p(x_i(t))\,\mc{CN}\big(z_{i}(t);x_i(t), 1/(\hat{\gamma}_{\da}(t)\lr{ \|\mb{h}_i\|^2})\big).
\end{align}
Thus, $q(x_i(t)) = p\big(x_i(t)\!\mid\! z_i(t) = x_i(t) + \mc{CN}\big(0,1/(\hat{\gamma}_{\da}(t)\lr{\|\mb{h}_i\|^2})\big)$. The variational  mean $\hat{x}_i(t)$ and variance $\tau_{x_i(t)}$ of $x_i(t)$ can be obtained from Corollary \ref{cor:meanvar}. Here, the matched-filter output is
\begin{align} \label{eq:z_t}
    z_{i}(t) &= \frac{\lr{\mb{h}_i^H}}{\lr{\|\mb{h}_i\|^2}} \bigg( \hat{\bs{\Phi}}(t)^H\mb{y}(t) - \sum_{j\neq i}^K \lr{\mb{h}_j} \lr{x_j(t)} \bigg) \nonumber\\
    & = \frac{\|\hat{\mb{h}}_i\|^2\hat x_i(t) + \hat{\mb{h}}_i^H \mb{r}(t)}{\lr{\|\mb{h}_i\|^2 }},
\end{align}
where $\lr{\|\mb{h}_i\|^2 }  = \|\hat{\mb{h}}_i\|^2 + \tr\{\bs{\Sigma}_{\mb{h}_i}\}$. 
\subsubsection{Extracting the Transmit Phase Noise $\theta_i(t)$}
Once the posterior update of $x_i(t)$ is obtained, we can extract the transmit PN at slot $t$, $\{\hat{\theta}_i(t), \kappa_{{\rm t},i}(t)\}$, and its predictive prior for slot $t+1$, $\{\hat{\theta}^p_i(t+1), \kappa_{{\rm t},i}^p(t+1)\}$ as follows.
The prior concentration $\varepsilon$ in Theorem~\ref{thm:main} is replaced by the predictive von Mises prior
\begin{equation}
\varepsilon_i^{p}(t) \triangleq \kappa_{{\rm t},i}^{p}(t)\, \e^{\ji\hat{\theta}_i^{p}(t)},
\label{eq:eps_p_def}
\end{equation}
carried forward from slot $t-1$ via the von Mises composition rule of Section~\ref{sec:preliminaries},
\begin{align}
\kappa_{{\rm t},i}^{p}(t) &= J^{-1}\!\big(J(\kappa_{{\rm t},i}(t-1))\, J(\varepsilon_{t,i})\big), \label{eq:kappa_predict}\\
\hat{\theta}_i^{p}(t) &= \hat{\theta}_i(t-1). \label{eq:theta_predict}
\end{align}
We have the observation precision $\gamma \triangleq \hat{\gamma}_\da(t)\langle\|\mb{h}_i\|^2\rangle$
and $z_i(t)$ derived in \eqref{eq:z_t}, the posterior natural parameter can be rewritten from \eqref{eq:eta} as
\begin{equation}
\eta_{k,i}(t) \triangleq \varepsilon_i^{p}(t) + 2\hat{\gamma}_\da(t)\langle\|\mb{h}_i\|^2\rangle\, z_i(t)\, s_k^*.
\label{eq:eta_ki}
\end{equation}
Theorem~\ref{thm:collapsedVM} (II) gives the discrete symbol posterior
\begin{equation}
\alpha_{k,i}(t) \propto p_k \exp\!\big(-\hat{\gamma}_\da(t)\langle\|\mb{h}_i\|^2\rangle |s_k|^2\big)\, I_0\!\big(|\eta_{k,i}(t)|\big).
\label{eq:alpha_ki}
\end{equation}
The exact transmit-phase posterior is then the finite von Mises mixture given by Theorem~\ref{thm:main} (III),
\begin{equation}
p\big(\theta_i(t)\,\big|\,z_i(t)\big) = \sum_k \alpha_{k,i}(t)\, \mc{VM}\big(\theta_i(t); \eta_{k,i}(t)\big),
\label{eq:theta_mixture_post}
\end{equation}
which is collapsed into a single von Mises distribution
$\mc{VM}\big(\theta_i(t); \hat{\theta}_i(t), \kappa_{{\rm t},i}(t)\big)$ via Theorem~\ref{thm:collapsedVM}:
\begin{align}
\hat{\theta}_i(t) &= \scalebox{1.5}{$\angle$}\!\left( \sum_k \alpha_{k,i}(t)\, J\big(|\eta_{k,i}(t)|\big)\, \e^{\ji\angle \eta_{k,i}(t)} \right), \label{eq:theta_hat_extract}\\[2pt]
\kappa_{{\rm t},i}(t) &= J^{-1}\!\left( \left| \sum_k \alpha_{k,i}(t)\, J\big(|\eta_{k,i}(t)|\big)\, \e^{\ji\angle \eta_{k,i}(t)} \right| \right). \label{eq:kappa_extract}
\end{align}
The pair $\{\hat{\theta}_i(t), \kappa_{{\rm t},i}(t)\}$ is then predicted forward to slot $t+1$ using the same rule as in \eqref{eq:kappa_predict}--\eqref{eq:theta_predict},
\begin{align}
\kappa_{{\rm t},i}^{p}(t+1) &= J^{-1}\!\big(J(\kappa_{{\rm t},i}(t))\, J(\varepsilon_{t,i})\big), \\
\hat{\theta}_i^{p}(t+1) &= \hat{\theta}_i(t),
\end{align}
which completes the recursion and initializes $\eta_{k,i}(t+1)$ in \eqref{eq:eta_ki} for the next slot.

\subsubsection{Update $\gamma_{\da}(t)$}
The variational distribution $q(\gamma_{\da}(t))$ can be obtained as 
\begin{align}  \label{eq:q_nu_da}
q(\gamma_{\da}(t)) &\propto \exp\big\{\blr{ \ln p({\mb{Y}_{\da}}|\mb{H},\{\bs{\phi}(t)\},\mb{X}_{\da}, \{\gamma_\da(t)\}) \nonumber\\
&\quad + \ln p(\{\gamma_\da(t)\}) }\big\} \nonumber\\
&\propto \exp\big\{(a_\da + M - 1)\ln \gamma_{\da}(t) \nonumber\\
&\quad  -\gamma_{\da}(t)(b_\da + \lr{\|\bs{\Phi}(t)^H\mb{y}(t) - \mb{H} \mb{x}(t) \|^2} )\big\}.
\end{align}
The mean of $q(\gamma_{\da}(t))$ is
\begin{align} \label{eq:gamma:data}
    \lr{\gamma_{\da}(t)} = \frac{a_\da +  M }{b_\da + \lr{\|\bs{\Phi}(t)^H\mb{y}(t) - \mb{H} \mb{x}(t) \|^2} },
\end{align}
where 
\begin{align} \label{eq:gamma:data:expand}
    &\lr{\|\bs{\Phi}(t)^H\mb{y}(t) - \mb{H} \mb{x}(t) \|^2} \nonumber \\
    &= \|\mb{r}(t)\|^2 + \sum_{m=1}^M|y_m(t)|^2(1-J(\eta_{\mr{r},m}(t))^2)  \nonumber\\
    &\quad + \sum_{i=1}^K \Big[\lr{|x_i(t)|^2} \tr\{\mb{\Sigma}_{\mb{h}_i} \} + \|\hat{\mb{h}}_i\|^2 \tau_{x_i(t)} \Big].
\end{align}

The MF-VB algorithm for data detection using the prior of $x_i(t)$ is summarized in Algorithm \ref{algo:VB-Detecting}.

\begin{algorithm}[!t]
	\small
	\caption{-- Joint Data Detection and Phase Noise Tracking}
	\label{algo:VB-Detecting}
	\begin{algorithmic}[1]
		\State \textbf{Input:} $\mb{y}(t)$, $\hat{\mb{H}}$, $\bs{\Sigma}_{\mb{h}_i}$, 
        $\{\hat\theta_i(T_\p), \kappa_{\mr{t},i}(T_\p)\}, \{\hat\phi_m(T_\p), \kappa_{\mr{r},m}(T_\p)\}$
		\State \textbf{Output:} $\hat{\mb{s}}$, $\hat{\bs{\theta}}$, and $\hat{\bs{\phi}}$\;
        \For{$i = 1,\ldots,K$} \Comment{Handoff from pilot}
        \State $\hat\theta^p_i(T_\p+1) \gets \hat\theta_i(T_\p)$
        \State $\kappa^p_{\mr{t},i}(T_\p+1)\gets J^{-1}\!\left(
		J\!\left(\kappa_{\mr{t},i}(T_\p)\right)\cdot J(\varepsilon_{\mr{t},i})
		\right)$
        \EndFor
        \For{$t = T_\p+1,\ldots,T_\p+T_\da$}
        \If{$t = T_\p+1$}
        \State $\bs{\phi}^p(t) \gets \hat{\bs{\phi}}(T_\p)$
        \State $\bs{\kappa}_{\mb{r}}^p(t) \gets \bs{\kappa}^p_{\mr{r}}(T_\p+1)$
        \Else
        \State $\bs{\phi}^p(t) \gets \hat{\bs{\phi}}(t-1)$
        \State $\bs{\kappa}_{\mb{r}}^p(t) \gets J^{-1} \!\left(
		J\!\left(\bs{\kappa}_{\mr{r}}(t-1)\right)\cdot J(\bs{\varepsilon}_{\mr{r}})\right)$
        \EndIf
        \For{$i = 1,\ldots,K$}
        \State $\varepsilon_i^{p}(t) \triangleq \kappa_{{\rm t},i}^{p}(t)\, \e^{\ji\hat{\theta}_i^{p}(t)}$,
        \EndFor
        \State Initialize $\hat{x}_i^0(t) = 0$, $\tau_{x_i(t)}^0 = \mr{Var}_{p(s_i(t))}[s_i(t)]$, $\forall i$.
        \State Initialize $\mb{r}(t) = \mb{y}(t) - \hat{\mb{H}}\hat{\mb{x}}^{0}(t)$, $\eta_{\mr{r},m}^0(t)\gets \kappa_{\mr{r},m}^p(t)\e^{\ji \hat\phi_m^p(t)}$ 
        \State Initialize $d_{\mr{r},m}^0(t) = {I_1( |\eta^0_{\mr{r},m}(t)|)}/{I_0(|\eta^0_{\mr{r},m}(t)|)}$,~$\forall m$
		\For{$\ell=1,2,\ldots$}
		\State Update $\hat{\gamma}^\ell_\da(t)$ via \eqref{eq:gamma:data} and \eqref{eq:gamma:data:expand}.
		\For{$i=1,2,\ldots,K$}
		\State $z_i^\ell(t) \gets ({\|\hat{\mb{h}}_i\|^2\hat x_i(t) + \hat{\mb{h}}_i^H \mb{r}(t)})/{\lr{\|\mb{h}_i\|^2 }}$
		\State $\phantom{z_i^\ell(t)} \mathllap{\hat{x}_i^{\ell+1}(t)} \gets \mathbb{E}\big[x_i(t)\!\mid\!z_i(t) = x_i(t)$
        \Statex \qquad\qquad \qquad\qquad\qquad $+  \,\,\mc{CN}\big(0,1/(\hat{\gamma}_\da(t)\lr{\|\mb{h}_i\|^2})\big)\big]$
		\State $\phantom{z_i^\ell(t)} \mathllap{\tau_{x_i(t)}^{\ell+1}} \gets \mr{Var}\big[x_i(t)\!\mid\!z_i(t)=x_i(t) $ 
        \Statex \qquad\qquad \qquad\qquad\qquad $+\,\,\mc{CN}\big(0,1/(\hat{\gamma}_\da(t)\lr{\|\mb{h}_i\|^2})\big)\big]$
        
		\State $\phantom{z_i^\ell(t)} \mathllap{\mb{r}(t)} \gets \mb{r}(t) + \hat{\mb{h}}_i (\hat{x}_i^\ell(t) - \hat{x}_i^{\ell+1}(t))$
    \EndFor
	\State $\bs{\eta}_{\mr{r}}(t) \gets 2\hat{\gamma}_\da^\ell(t) \big(\mb{y}(t)\odot\mb{y}^*(t)\odot\mb{d}_{\mr{r}}(t)\odot \e^{\ji\hat{\bs{\phi}}^\ell(t)}$
    \Statex \qquad\qquad \qquad\qquad\qquad$ - \mb{y}(t)\odot\mb{r}^*(t)\big)$
	\State $\phantom{\bs{\eta}_{\mr{r}}} \mathllap{\hat{\bs{\phi}}^{\ell+1}}(t) \gets \angle(\bs{\kappa}^p_{\mr{r}}(t)\, \e^{\ji\hat{\bs{\phi}}^p(t)} + \bs{\eta}_{\mr{r}}(t))$ 
	\State $\phantom{\bs{\eta}_{\mr{r}}} \mathllap{\bs{\kappa}_{\mr{r}}^{\ell+1}}(t) \gets |\bs{\kappa}^p_{\mr{r}}(t)\, \e^{\ji\hat{\bs{\phi}}^p(t)} + \bs{\eta}_{\mr{r}}(t)|$
	\State $\phantom{\bs{\eta}_{\mr{r}}} \mathllap{\mb{d}_{\mr{r}}^{\ell+1}}(t) \gets {I_1(\bs{\kappa}_{\mr{r}}^{\ell+1}(t))}./{I_0(\bs{\kappa}_{\mr{r}}^{\ell+1}(t))}$
	\State $\phantom{\bs{\nu}_{\mr{r}}} \mathllap{\mb{r}}(t) \gets \mb{r}(t) -  \mb{y}(t)\odot \big( \mb{d}_{\mr{r}}^{\ell}(t)\odot\e^{-\ji \hat{\bs{\phi}}^{\ell}(t)} $
    \Statex \qquad\qquad \qquad\qquad\qquad$- \mb{d}_{\mr{r}}^{\ell+1}(t)\odot\e^{-\ji \hat{\bs{\phi}}^{\ell+1}(t)} \big)$\;
    \For{$i=1,\ldots,K$} \Comment{Extract transmit PN}
    \State $\hat{\theta}_i(t) = \angle\!\left( \sum_k \alpha_{k,i}(t)\, J\big(|\eta_{k,i}(t)|\big)\, \e^{\ji\angle \eta_{k,i}(t)} \right)$, 
    \State $\kappa_{{\rm t},i}(t) = J^{-1}\!\Big( \Big| \sum_k \alpha_{k,i}(t)\, J\big(|\eta_{k,i}(t)|\Big)\, \e^{\ji\angle \eta_{k,i}(t)} \Big| \Big).$ 
    \State $\kappa_{{\rm t},i}^{p}(t+1) = J^{-1}\!\big(J(\kappa_{{\rm t},i}(t))\, J(\varepsilon_{t,i})\big)$ \\
    \State $\hat{\theta}_i^{p}(t+1) = \hat{\theta}_i(t),$
    \EndFor
	\EndFor
    \EndFor
	\State $\forall i: \hat{s}_i(t) \leftarrow \arg \max_{a\in \mathcal{A}} p(a\!\mid\!z_i(t))$ using \eqref{eq:ps} of Theorem \ref{thm:main}\;
\end{algorithmic}
\end{algorithm}

\section{Channel Refinement via Virtual Pilots} \label{sec:ChannelRefinement}
The VB-Detecting approach in Section \ref{sec:DataDetection} treats $\hat{\mb{H}}$ as fixed for the entire data phase. 
Although the physical channel is assumed to be block-fading, the evolving PN causes the effective channel observed by the receiver to vary over the data frame. We call this effect PN-induced channel aging. For long data frames, this effect can cause the mismatch for the fixed channel estimate, motivating periodic channel refinement using reliably detected data symbols as virtual pilots. We now extend the VB framework to periodically re-estimate $\mb{H}$ during the data phase, referred to as VB-Refining, using the soft posterior moments of $x_i(t)$ from Corollary \ref{cor:meanvar} as virtual pilots. We divide the data phase into $N_B$ blocks of length $L$, with $T_\da = N_B L$:
\begin{align}
    \mc{B}_n = [T_\p+(n-1)L+1, T_\p + nL], \, n = 1,\ldots, N_B.
\end{align}
Within block $\mc{B}_n$, Algorithm \ref{algo:VB-Detecting} runs with the channel posterior fixed at $q^{(n-1)}(\mb{h}_i) = \mc{CN}(\hat{\mb{h}}_i^{(n-1)}, \bs{\Sigma}_{\mb{h}_i}^{(n-1)})$. At the end of block $\mc{B}_n$, a channel refinement step produces an updated posterior $q^{(n)}(\mb{h}_i)$, which becomes the prior for block $\mc{B}_{n+1}$. This requires reopening the mean-field factorization in \eqref{eq:mean-field:channel} to let $q(\mb{h}_i)$ remain a live factor throughout the data phase. The local factorization for block $\mc{B}_{n}$ is
\begin{align} \label{eq:mean-field:refinement}
		&p(\mb{H},\{\bs{\phi}(t)\}_{t=1}^{T_\p},\{\bs{\theta}(t)\}_{t=1}^{T_\p},\gamma_\p, \{\mb{x}(t)\}_{t\in \mc{B}_n}, \{\bs{\phi}(t)\}_{t\in \mc{B}_n}, \nonumber\\
        &\qquad \{\gamma_\da(t)\}_{t\in \mc{B}_n} \!\mid\! \mb{Y}_\p, \mb{Y}_{\mc{B}_n};\mb{S}_\p)\nonumber\\
		&= \Bigg[\prod_{i=1}^K q^{(n)}(\mb{h}_i)\Bigg] q(\gamma_\p)  \Bigg[\prod_{t=2}^{T_\p} \prod_{i=1}^K q(\theta_i(t))  \prod_{m=1}^M q(\phi_m(t))\Bigg]\nonumber\\
        &\times \Bigg[\prod_{t\in \mc{B}_n}\prod_{i=1}^K q(x_i(t)) \prod_{m=1}^M q(\phi_m(t)) q(\gamma_\da(t))\Bigg].
\end{align}
To prevent the error propagation from unreliable decisions, we restrict the channel update to only the slots where the discrete symbol posterior $\alpha_k(t)$ from Theorem 1 is sharply peaked on one constellation point:
\begin{align} \label{eq:th}
\tilde{\mc{B}}_{n,i} = \Big\{t \in \mc{B}_{n}:\max_{i} \alpha_{k,i}(t) > \rho
\Big\},
\end{align}
where $\rho$ is the threshold. The channel variational posterior is then derived using only $t \in \tilde{\mc{B}}_{n,i}$ as follows:

\emph{Update $\mb{h}_i$:} Taking the expectation of the conditional \eqref{eq:con} w.r.t. all latent variables except $\mb{h}_i$, and using $q^{(n-1)}(\mb{h}_i) = \mc{CN}(\hat{\mb{h}}_i^{(n-1)}, \bs{\Sigma}_{\mb{h}_i}^{(n-1)})$ as the prior in place of $p(\mb{h}_i,\mb{C}_i)$, $q(\mb{h}_i)$ can be derived as
\begin{align}  \label{eq:h:0}
    q(\mb{h}_i)\propto &\exp\bigg\{\bblr{ \sum_{t \in \tilde{\mc{B}}_n} \ln p\big(\mb{y}(t)\!\mid\!\mb{H},  \bs{\phi}(t), \gamma_\da(t),\mb{x}(t) \big) \nonumber\\
    &\qquad + \ln q^{(n-1)}(\mb{h}_i)}_{-\mb{h}_i} \bigg\}
\end{align}
which, following the same expansion as \eqref{eq:h:CE}, yields the variational covariance and mean as
\begin{align}
\bs{\Sigma}_{\mb{h}_i}^{(n)} &= \bigg[\hat{\gamma}_\da \sum_{t\in \tilde{\mc{B}}_n}  \lr{|x_i(t)|^2} \mb{I}_M + \big(\bs{\Sigma}_{\mb{h}_i}^{(n-1)}\big)^{-1} \bigg]^{-1}, \\
\hat{\mb{h}}_i^{(n)} &= \bs{\Sigma}_{\mb{h}_i}^{(n)} \Big[\big(\bs{\Sigma}_{\mb{h}_i}^{(n-1)}\big)^{-1}\hat{\mb{h}}_i^{(n-1)} \nonumber\\
&\quad + \hat{\gamma}_\da \sum_{t \in \tilde{\mc{B}}_n}\big(\mb{r}(t) +\hat{\mb{h}}_i^{(n-1)} \hat{x}_i(t) \big)\hat{x}_i^*(t) \Big]. \label{eq:h:mean:refined}
\end{align}
The VB-Refining procedure is summarized in Algorithm \ref{algo:VB-Refining}.

\begin{algorithm}
\small
\caption{-- Channel Refinement and Phase Noise Tracking}
\label{algo:VB-Refining}
\begin{algorithmic}[1]
\State \textbf{Input:} $\mb{y}(t)$, with $t = T_\p+1,\ldots,T_\p+T_\da$; 
\State Initialize $\hat{\mb{h}}_i^{(0)} \gets \hat{\mb{h}}_i$, $\bs{\Sigma}_{\mb{h}_i}^{(0)} \gets \bs{\Sigma}_{\mb{h}_i}$ from \textbf{Algorithm~\ref{algo:VB-Tracking}}; priors $\{p(s_i(t))\}$; terminal phase states $\{\hat\theta_i(T_\p),\kappa_{\mr{t},i}(T_\p)\}$, $\{\hat\phi_m(T_\p),\kappa_{\mr{r},m}(T_\p)\}$; block length $L$; threshold $\rho$
\State \textbf{Output:} $\hat{\mb{s}}$, $\hat{\bs\theta}$, $\hat{\bs\phi}$, $\{\hat{\mb{h}}_i^{(N_B)}, \bs{\Sigma}_{\mb{h}_i}^{(N_B)}\}$
\State Initialize $N_B \gets T_\da / L$
\State Initialize $\hat\theta_i^p(T_\p+1) \gets \hat\theta_i(T_\p)$, $\hat\phi_m^p(T_\p+1) \gets \hat\phi_m(T_\p)$
\State Initialize $\kappa_{\mr{t},i}^p(T_\p+1) \gets J^{-1}(J(\kappa_{\mr{t},i}(T_\p))\cdot J(\varepsilon_{\mr{t},i}))$, $\forall i$ 
\State Initialize $\kappa_{\mr{r},m}^p(T_\p+1) {\gets} J^{-1}(J(\kappa_{\mr{r},m}(T_\p))\cdot J(\varepsilon_{\mr{r},m}))$, $\forall m$ 
\For{$n = 1, \ldots, N_B$}
    \State $\mathcal{B}_n \gets [T_\p+(n-1)L+1,\, T_\p+nL]$
    \State $\hat{\mb{H}}^{(n-1)} \gets [\hat{\mb{h}}_1^{(n-1)}, \ldots, \hat{\mb{h}}_K^{(n-1)}]$
    \For{$t \in \mathcal{B}_n$} \hfill $\triangleright$ Run Algorithm~\ref{algo:VB-Detecting} with $\hat{\mb{H}}^{(n-1)}$
        \State $[\hat{x}_i(t), \tau_{x_i}(t), \{\alpha_{k,i}(t)\}, \hat\theta_i(t), \hat\phi_m(t), \kappa_{\mr{t},i}(t), \kappa_{\mr{r},m}(t)]$
        \Statex \qquad\qquad\quad $ \gets$ \textbf{Algorithm~\ref{algo:VB-Detecting}}$(\mathbf{y}(t), \hat{\mb{H}}^{(n-1)}, \bs{\Sigma}_{\mb{h}_i}^{(n-1)})$
        \State $\alpha_{k^*,i}(t) \gets \max\limits_k \alpha_{k,i}(t)$, $\forall i$
    \EndFor
    \State \textbf{Channel refinement:}
    \State $\tilde{\mc{B}}_{n,i} \gets \{t \in \mc{B}_n : \alpha_{k^*,i}(t) > \rho\}$ 
    \For{$i = 1, \ldots, K$}
        \State $E_i \gets \sum_{t \in \tilde{\mc{B}}_{n,i}} \big(|\hat{x}_i(t)|^2 + \tau_{x_i}(t)\big)$ 
        \State $\bs{\Sigma}_{\mb{h}_i}^{(n)} \gets \big[\hat\gamma_\da E_i\, \mb{I}_M + (\bs{\Sigma}_{\mb{h}_i}^{(n-1)})^{-1}\big]^{-1}$ 
        \State $\hat{\mb{h}}_i^{(n)} \gets \bs{\Sigma}_{\mb{h}_i}^{(n)}\Big[(\bs{\Sigma}_{\mb{h}_i}^{(n-1)})^{-1}\hat{\mb{h}}_i^{(n-1)} $
        \Statex \qquad\qquad\qquad $+ \hat\gamma_\da \sum_{t \in \tilde{\mc{B}}_{n,i}} \big(\mb{r}(t) + \hat{\mb{h}}_i^{(n-1)}\hat{x}_i(t)\big)\hat{x}_i^*(t)\Big]$ 
        \For{$t \in {\mc{B}}_n$}
            \State $\mb{r}(t) \gets \mb{r}(t) + \big(\hat{\mb{h}}_i^{(n-1)} - \hat{\mb{h}}_i^{(n)}\big)\hat{x}_i(t)$ 
        \EndFor
    \EndFor
    \State $\hat\theta_i^p(T_\p+nL+1) \gets \hat\theta_i(T_\p+nL)$
    \State $\hat\phi_m^p(T_\p+nL+1) \gets \hat\phi_m(T_\p+nL)$
    \State $\kappa_{\mr{t},i}^p(T_\p+nL+1) \gets J^{-1}(J(\kappa_{\mr{t},i}(T_\p+nL))\cdot J(\varepsilon_{\mr{t},i}))$
    \State $\kappa_{\mr{r},m}^p(T_\p+nL+1) {\gets} J^{-1}(J(\kappa_{\mr{r},m}(T_\p+nL))\cdot J(\varepsilon_{\mr{r},m}))$
    \State $\hat{\mb{H}}^{(n)} \gets [\hat{\mb{h}}_1^{(n)}, \ldots, \hat{\mb{h}}_K^{(n)}]$
\EndFor
\State $\forall i, t$: $\hat{s}_i(t) \gets \arg\max\limits_{a \in \mathcal{A}} p(a | z_i(t))$ using (II) of Theorem 1
\end{algorithmic}
\end{algorithm}

\subsection{Complexity Analysis}
\label{sec:complexity}
Let $N=|\mathcal{A}|$ denote the constellation size, $L_\p$ and $L_\da$ the CAVI iteration numbers for VB-Tracking and VB-Detecting, respectively, and $N_B=T_\da/L$ the number of refinement blocks. We consider i.i.d. Rayleigh fading, i.e.,
$\mb{C}_i=\mb{I}_M/M$, so that $\bs{\Sigma}_{\mb{h}_i}$ in \eqref{eq:h:mean} and
$\bs{\Sigma}_{\mb{h}_i}^{(n)}$ in \eqref{eq:h:mean:refined} are scalar multiples of $\mb{I}_M$. Every channel-posterior update is then a vector operation of cost $\mathcal{O}(M)$, and
$\mathrm{Tr}\{\bs{\Sigma}_{\mb{h}_i}\}$ is obtained in $\mathcal{O}(1)$. The resulting
computational complexities are summarized in Table~\ref{tab:complexity}.

For VB-Tracking (Algorithm~\ref{algo:VB-Tracking}), each CAVI iteration consists of the
noise-precision update, the channel update for $K$ users, and the transmit- and
receive-PN updates over $T_\p$ slots. Each of these costs $\mathcal{O}(KMT_\p)$ or less,
so the per-iteration cost is $\mathcal{O}(KMT_\p)$, yielding
$\mathcal{O}(L_{\mathrm p}KMT_\p)$ overall.

VB-Detecting (Algorithm~\ref{algo:VB-Detecting}) requires $\mathcal{O}(K(M+N))$ per CAVI
iteration and slot, accounting for the residual/phase updates and the $N$-point discrete
data posterior. Thus, its total complexity is $\mathcal{O}(L_\da T_\da K(M+N))$.

VB-Refining (Algorithm~\ref{algo:VB-Refining}) has the same detection cost. In each block
and for each user, the virtual-pilot sum in (60) and the residual correction cost
$\mathcal{O}(ML)$, and the covariance and mean updates cost $\mathcal{O}(M)$. Summing
over $N_B$ blocks with $N_BL=T_\da$, the refinement overhead is
$\mathcal{O}(T_\da KM)$, which does not depend on $N_B$
because the blocks partition the data slots. The total complexity is
$\mathcal{O}\big(L_\da T_\da K(M+N)+T_\da KM\big)$, which has the same order as
VB-Detecting for $L_\da\geq1$.

For the baselines, EKF-EP \cite{wang2025iterative} requires
$\mathcal{O}(M+K+L_{\mathrm{EP}}KN)$ per slot and $\mathcal{O}(K^2M)$ per refinement
block, so its refinement cost grows with $N_B$. EKF-ZF \cite{reggiani2018extended}
requires $\mathcal{O}(M+K+KM)$ per slot, plus a one-time $\mathcal{O}(K^2M+K^3)$
pseudo-inverse computation. JED \cite{nguyen2024variational} has the same asymptotic
order as VB-Detecting, $\mathcal{O}(L_{\mathrm{JED}}T_\da K(M+N))$, but with fewer
phase-related operations.

Overall, the proposed VB approaches scale linearly with $M$, $K$, $N$, and the number of iterations, while avoiding the higher-order matrix operations required by the EKF baselines.

\begin{table}[t]
\centering
\caption{Total computational complexity of the proposed and baseline algorithms
(i.i.d.\ channels).}
\label{tab:complexity}
\begin{tabular}{@{}ll@{}}
\toprule
\textbf{Algorithm} & \textbf{Total complexity} \\
\midrule
VB-Tracking (Alg.~1)  & $\mathcal{O}(L_{\mathrm p}KMT_{\mathrm p})$ \\
VB-Detecting (Alg.~2) & $\mathcal{O}(L_\da T_\da K(M+N))$ \\
VB-Refining (Alg.~3)  & $\mathcal{O}\!\big(L_\da T_\da K(M+N) + T_\da KM\big)$ \\
\addlinespace
EKF-EP \cite{wang2025iterative} & $\mathcal{O}\!\big(T_\da (M+K+L_{\mathrm{EP}}KN) + N_BK^2M\big)$ \\
EKF-ZF \cite{reggiani2018extended} & $\mathcal{O}\!\big(T_\da (M+K+KM) + K^2M + K^3\big)$ \\
JED \cite{nguyen2024variational} & $\mathcal{O}(L_{\mathrm{JED}}T_\da K(M+N))$ \\
\bottomrule
\end{tabular}
\end{table}

\section{Simulation Results}
In this section, we evaluate the proposed VB-Tracking, VB-Detecting, and VB-Refining, presented in Algorithms~\ref{algo:VB-Tracking}, \ref{algo:VB-Detecting}, and \ref{algo:VB-Refining}, respectively, via Monte Carlo simulations. We consider an uplink massive MIMO system with $M = 40$ receive antennas at the BS and $K = 8$ single-antenna users, using orthogonal training sequences and a pilot length of $T_\p = 16$. We set $\rho$ in \eqref{eq:th} as $0.95$. The noise variance $N_0$ is set based on the SNR, which is defined as
\begin{align}
    {\rm SNR} = \frac{\mathbb{E}[\|\bs{\Phi}\mb{Hx}\|^2]}{\mathbb{E}[\|\mb{n}\|^2]} =\frac{\sum_{i=1}^K \tr\{\mb{C}_i\}}{MN_0}= \frac{K}{M N_0}.
\end{align}
We consider both i.i.d.\ Rayleigh fading, i.e., $\mb{C}_i=(1/M)\mb{I}_M$, and spatially correlated Rayleigh fading, where each column of $\mb{H}$ follows an exponential correlation model with covariance $\mb{C}_i$ given by
\begin{equation}
[\mb{C}_i]_{mk} =
\begin{cases}
(1/M)\alpha^{\,m-k}, & m \geq k, \\[2pt]
(1/M)\big(\alpha^{\,k-m}\big)^{*}, & m < k,
\end{cases}
\label{eq:corr_model}
\end{equation}
where $\alpha \in \mathbb{C}$, $|\alpha| \leq 1$, controls the correlation strength and phase between antenna elements. The true channel is generated as $\mb{h}_i \sim \mathcal{CN}(\mb{0}, \mb{C}_i)$. The transmit and receive phase noise processes follow the Wiener model in \eqref{eq:Wiener} with standard deviations $\sigma_{\mr{t},i} = \sigma_{\mr{r},m} = 3^\circ$ per slot, corresponding to von Mises concentration parameters $\varepsilon_{\mr{t},i} = 1/\sigma_{\mr{t},i}^2$ and $\varepsilon_{\mr{r},m} = 1/\sigma_{\mr{r},m}^2$. Following the reference phase convention in Section~\ref{subsubsec:Identifiability}, the initial phases $\phi_m(1)$ and $\theta_i(1)$ are absorbed into an effective channel $\tilde{\mb{h}}_i \triangleq \mr{diag}(\e^{\ji\bs{\phi}(1)}) \mathbf{h}_i \e^{\ji\theta_i(1)}$, with $\phi_1(1) = 0$ fixed as the global phase reference. All channel estimates $\hat{\mb{h}}_i$ are evaluated against $\tilde{\mb{h}}_i$, and all phase noise estimates $\hat\phi_m(t)$, $\hat\theta_i(t)$ are evaluated relative to their respective $t=1$ values, as defined below.
The NMSE of the estimated channel is defined as
\begin{align}
\mr{NMSE}({\mb{H}}) = {\big\| \hat{\mb{H}} - \tilde{\mb{H}} \big\|_F^2}/{\big\| \tilde{\mb{H}} \big\|_F^2},
\label{eq:nmse_channel}
\end{align}
where $\hat{\mb{H}} = [\hat{\mb{h}}_1, \ldots, \hat{\mb{h}}_K]$ is the estimated channel matrix, and $\tilde{\mb{H}}$ is the corresponding effective channel.
We compare our VB algorithms with the following baselines:
\begin{itemize}
    \item EKF-EP \cite{wang2025iterative}: The data-aided EKF approach is used for phase noise estimation, and the channel is re-estimated using data symbols as virtual pilots. The expectation propagation (EP) is used for data detection.
    \item JED \cite{nguyen2024variational}: joint channel estimation and data detection using a VB framework. In this JED, the PN is absorbed into the channel to estimate the channel.
    \item EKF-ZF \cite{reggiani2018extended}: This method uses an extended Kalman filter for PN tracking and zero-forcing for data detection. Since it addresses only data detection, we feed the channel estimated by our VB-Tracking algorithm into the EKF-ZF for detection.
   \item {Genie-CE: The true phase noises $\bs{\theta}(t)$ and $\bs{\phi}(t)$ are assumed to be known during the pilot transmission phase. Thus, we use Genie-CE with the LMMSE channel estimator as the lower bound for the channel NMSE.}
\end{itemize}

\begin{figure}
  	\centering
 	\includegraphics[width=1.0\linewidth]{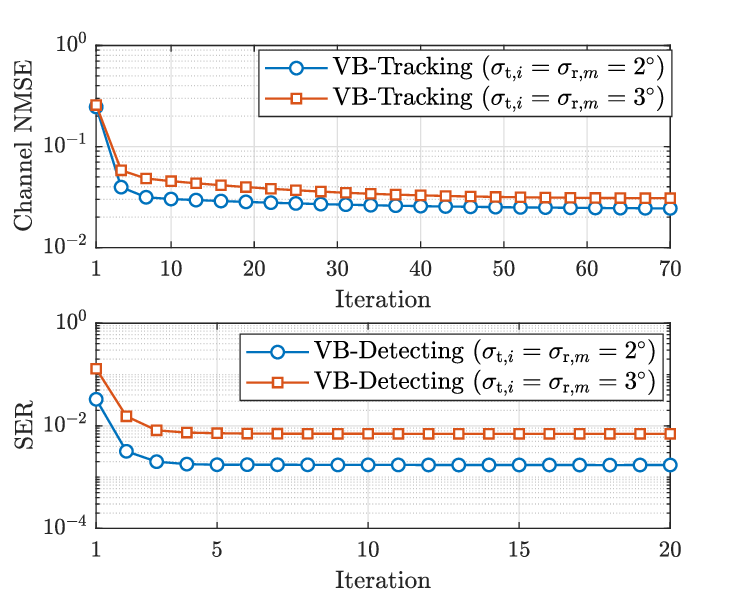}
 	\caption{Convergence of VB-Tracking (Alg. 1) and VB-Detecting (Alg. 2), assuming i.i.d. fading channels with $M=40$, $K = 8$, $T_\p=16$, $\mr{SNR} = 15$ dB, and $T_\da = 100$.}
 	\label{NMSE_SER_Con}
\end{figure}
We first show the convergence behaviors of the proposed VB algorithms as in Fig.~\ref{NMSE_SER_Con}. The channel-estimation VB-Tracking and the data-detection VB-Detecting converge rapidly, with channel NMSE and SER stabilizing within about 20 and 5 iterations, respectively. This fast convergence reflects the effectiveness of derived closed-form expressions of variational distributions and coordinate-ascent updates enabled by the von Mises conjugacy.

\begin{figure}
  	\centering
 	\includegraphics[width=1.0\linewidth]{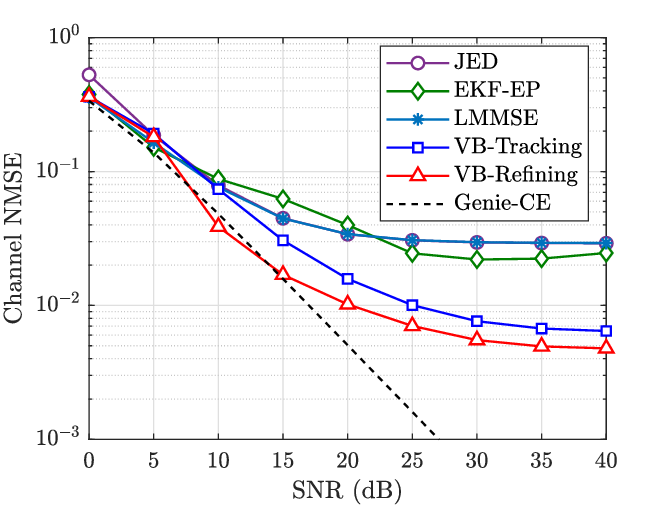}
 	\caption{NMSE vs. SNR, assuming i.i.d. fading channels with $M=40$, $K = 8$, $T_\p=16$, $\sigma_{\mr{t},i} = \sigma_{\mr{r},m} = 3^\circ$, $T_\da = 100$, and 16-QAM signaling.}
 	\label{NMSE_SNR_new}
\end{figure}
In Fig. \ref{NMSE_SNR_new}, we show the NMSE as a function of SNR, assuming i.i.d. channels. When SNR increases, VB-Refining separates from JED and EKF-EP and tracks closer to the Genie-CE bound, but its NMSE flattens rather than continuing to fall linearly with SNR since even perfect knowledge of the noise level cannot eliminate the residual uncertainty in tracking a continuously drifting Wiener phase process.


\begin{figure*}[!t]
    \centering
    \subfloat[NMSE vs. SNR, assuming correlated fading channels.]{\includegraphics[width=0.32\linewidth]{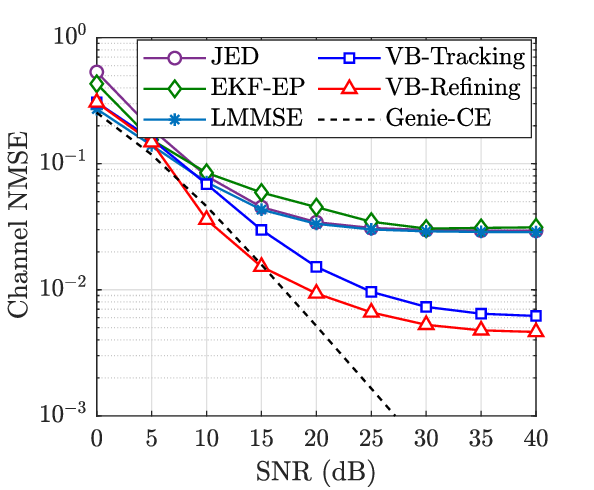}
        \label{NMSE_User}}
    \hfill
    \subfloat[NMSE vs. PN standard deviation.]{
    \includegraphics[width=0.32\linewidth]{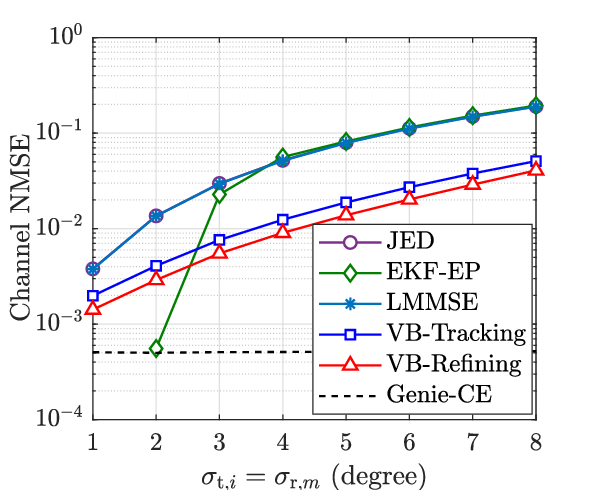}
    \label{SER_Tdata}}
    \hfill
    \subfloat[NMSE of VB-Refining after each refinement block.]{
    \includegraphics[width=0.32\linewidth]{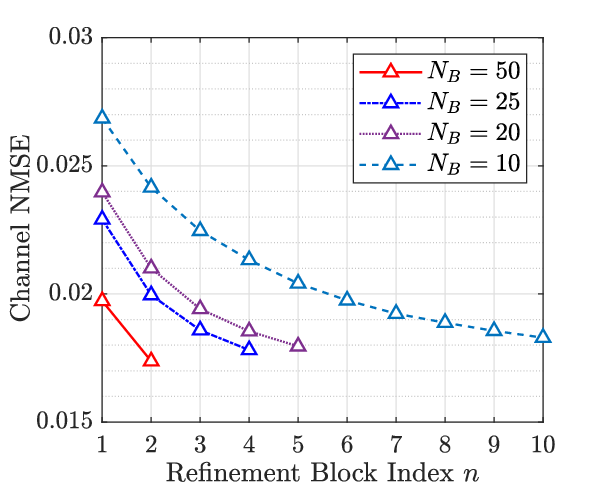}
    \label{BLER_Td}}
    \caption{Channel NMSE performance under different PN and channel conditions, with $M=40$, $K=8$, $T_\p=16$, $T_\da=100$, and 16-QAM signaling. (a) Assuming correlated fading channels with
$\sigma_{\mr{t},i}=\sigma_{\mr{r},m}=3^\circ$;
    (b) Assuming i.i.d. fading channels with $\mr{SNR}=30$ dB; and
    (c) Assuming i.i.d. fading channels with
$\sigma_{\mr{t},i}=\sigma_{\mr{r},m}=3^\circ$.}
    \label{fig:NMSE_results}
\end{figure*}

The qualitative trends in Fig. \ref{fig:NMSE_results}(a) are consistent with Fig. \ref{NMSE_SNR_new}, with VB-Refining again achieving the lowest NMSE. The channel NMSE values are close to the i.i.d. case, indicating that the proposed VB framework is robust to the spatial correlation typical of mmWave and massive MIMO deployments.

Fig. \ref{fig:NMSE_results}(b) shows the NMSE versus the standard deviation of PN. Channel NMSE increases monotonically with PN standard deviation across all methods, as expected since larger phase innovations make the Wiener process harder to track and corrupt the effective channel more severely. VB-Refining consistently outperforms VB-Tracking, JED, and EKF-EP, with the gap widening at higher phase-noise levels, showing that periodic re-estimation using virtual pilots is especially valuable when phase drift is severe. The Genie-CE bound stays far below all algorithms, confirming that phase-noise-induced uncertainty is the dominant error floor in this regime.

In Fig. \ref{fig:NMSE_results}(c),  channel NMSE decreases with the refinement block index for all $L\in\{50,25,20,10\}$ ($N_B=T_\da/L\in\{2,4,5,10\}$), showing that successive virtual-pilot updates extract useful channel information from detected data. Less frequent refinement, i.e., larger $N_B$, starts with longer blocks and therefore achieves a lower NMSE earlier: $L=50$ reaches $\approx0.0175$ after two blocks, while $L=10$ requires all ten blocks to reach $\approx 0.0185$. The final NMSE gap is only $\sim10^{-3}$, indicating that frequent refinement is unnecessary. Moreover, the proposed VB-Refining has the refinement cost $\mathcal{O}(T_\da KM)$ independent of $N_B$, whereas EKF-EP costs $\mathcal{O}(N_BK^2M)$. Thus, coarser schedules reduce posterior updates, signaling, and latency without increasing VB-Refining's arithmetic cost; for $T_\da=100$, $L=50$ captures most of the NMSE gain with only two refinements.

\begin{figure}
  	\centering
 	\includegraphics[width=1.0\linewidth]{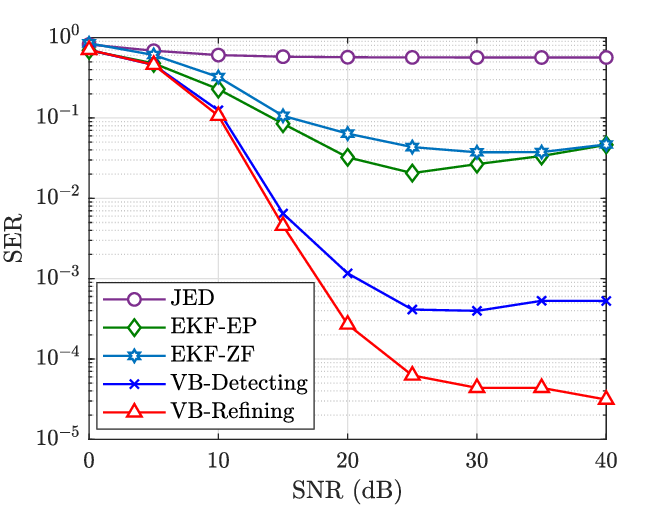}
 	\caption{SER vs. SNR, assuming i.i.d. fading channels with $M=40$, $K = 8$, $T_\p=16$, $\sigma_{\mr{t},i} = \sigma_{\mr{r},m} = 3^\circ$, $T_\da = 100$, and 16-QAM signaling.}
 	\label{SER_SNR_new}
\end{figure}
Fig. \ref{SER_SNR_new} shows the SER versus SNR, assuming i.i.d. channels. VB-Refining and VB-Detecting achieve substantially lower SER than JED, EKF-EP, and EKF-ZF baselines, with JED flooring around SER = 0.5, confirming that absorbing PN directly into the channel estimate as JED does is inadequate once PN evolves during the data phase. The advantage of VB-Refining over VB-Tracking grows at high SNR. As thermal noise is no longer the bottleneck, channel aging from holding $\hat{\mathbf{H}}$ fixed after the pilot phase becomes the main source of residual error. VB-Refining corrects this channel aging by periodically refining the estimate using reliable virtual pilots.

\begin{figure}
  	\centering
 	\includegraphics[width=1.0\linewidth]{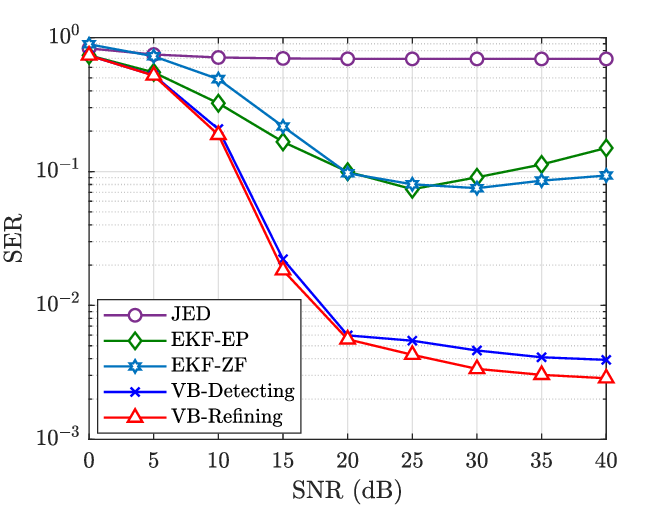}
 	\caption{SER vs. SNR, assuming correlated fading channels with $M=40$, $K = 8$, $T_\p=16$, $\sigma_{\mr{t},i} = \sigma_{\mr{r},m} = 3^\circ$, $T_\da = 100$, and 16-QAM signaling.}
 	\label{SER_SNR_non_iid}
\end{figure}
We illustrate the SER versus SNR assuming correlated channels in Fig. \ref{SER_SNR_non_iid}. The EKF-EP and EKF-ZF algorithms show a non-monotonic SER: degrading at moderate SNR before improving again; this could be due to the Kalman linearization process becoming less accurate under correlated fading interference, whereas the VB methods degrade gracefully and retain their advantage throughout the process. VB-Refining achieves the lowest SER among the evaluated methods and closely follows VB-Detecting, indicating that the proposed VB framework remains effective for data detection under spatially correlated channels. This further supports that the CAVI-based updates handle inter-user interference structure more reliably than EKF-linearized approaches.

\begin{figure}
  	\centering
 	\includegraphics[width=1.0\linewidth]{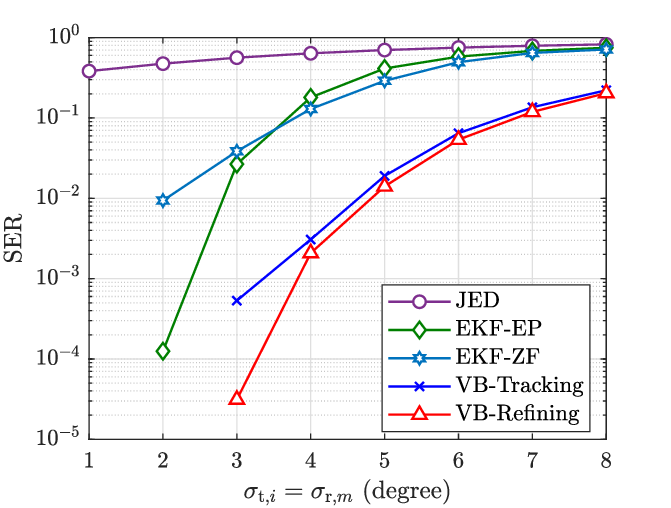}
 	\caption{SER vs. standard deviation of PNs, assuming i.i.d. fading channels with $M=40$, $K = 8$, $T_\p=16$, $T_\da = 100$, $\mr{SNR} = 30$ dB, and 16-QAM signaling.}
 	\label{SER_Std}
\end{figure}
Fig. \ref{SER_Std} shows the SER versus standard deviation of PN. The SER for all algorithms deteriorates sharply as phase noise std increases beyond $3^\circ-4^\circ$, but VB-Refining and VB-Detecting maintain a lower error floor than the EKF and JED baselines even at $8^\circ$ innovation std. JED again performs poorly across the entire range, reaffirming that separating the phase-noise and channel estimation problems is critical for robustness. The convergence of all curves at very high phase-noise levels suggests a fundamental performance limit where phase uncertainty overwhelms any algorithmic advantage.

Since the terminal phase noise estimates $\hat\phi_m(T_\p)$ and $\hat\theta_i(T_\p)$ are propagated forward as the predictive priors for data detection, we evaluate phase tracking accuracy at $t = T_\p$. Since $\hat\phi_m(t)$ and $\hat\theta_i(t)$ track the phase evolution relative to $t=1$ under the absorption convention, we define the relative ground-truth trajectories $\phi_m^{\mr{true}}(t) \triangleq \phi_m(t) - \phi_m(1), 
\theta_i^{\mr{true}}(t) \triangleq \theta_i(t) - \theta_i(1)$, then the wrapped (circular) phase errors at any time $t$ are
$\epsilon_{\phi,m}(t) \!\triangleq \angle\big( \e^{\ji(\hat\phi_m(t) - \phi_m^{\mr{true}}(t))} \big), 
\epsilon_{\theta,i}(t) \triangleq \angle\big( \e^{\ji(\hat\theta_i(t) - \theta_i^{\mr{true}}(t))} \big).$
Thus, the PN MSE at $t$ at the BS and users are expressed as
\begin{align}
\mr{MSE}_{\phi}(t) =  \frac{1}{M} \!\sum_{m=1}^{M}\! \epsilon_{\phi,m}^2(t), \,\,
\mr{MSE}_{\theta}(t) =  \frac{1}{K} \!\sum_{i=1}^{K}\! \epsilon_{\theta,i}^2(t).
\end{align}

\begin{figure}
  	\centering
 	\includegraphics[width=1.0\linewidth]{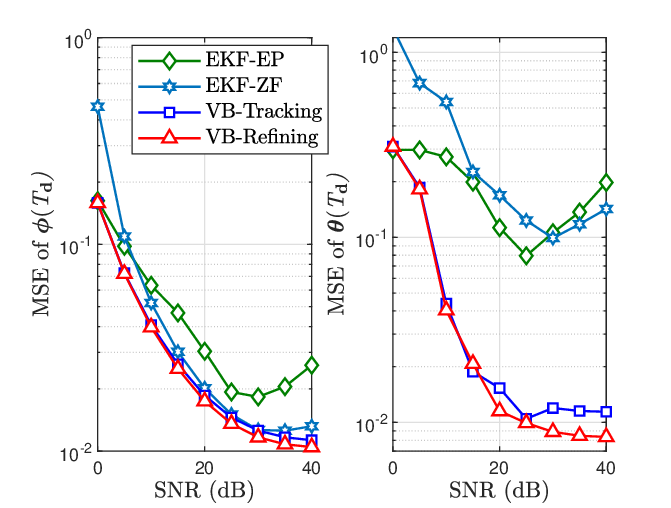}
 	\caption{MSE of PNs at $T_\da$, assuming i.i.d. fading channels with $M=40$, $K = 8$, $T_\p=16$, $T_\da = 100$, $\mr{SNR} = 20$ dB, and 16-QAM signaling.}
 	\label{MSE_PN}
\end{figure}
In Fig. \ref{MSE_PN}, we show the MSE of phase noise estimates at $T_\da$. VB-Tracking and VB-Refining achieve the lowest phase MSE for both receive and transmit PNs across most of the SNR range, with EKF-EP showing a distinctive non-monotonic dip-then-rise pattern for both $\theta$ and $\phi$ estimation, again pointing to linearization sensitivity under interference. The MSE plateaus at high SNR for the VB algorithms, indicating that PN tracking error becomes dominated by the phase-noise innovation variance itself rather than the residual channel-estimation noise. This SNR-saturation behavior parallels the channel NMSE floor in Fig. \ref{NMSE_SNR_new}, jointly confirming that phase noise, not thermal noise, is the fundamental bottleneck the proposed VB framework is designed to mitigate.

\section{Conclusion}
This paper has developed a novel VB framework for joint channel estimation, data detection, and PN tracking in massive MIMO systems. By leveraging von Mises priors, VB-Tracking derives closed-form CAVI updates during the pilot phase without requiring the Jacobian linearization used in EKF-based approaches. VB-Detecting then absorbs transmit PN into the signal under a sequentially propagated von Mises predictive and achieves an exact posterior of that composite variable. VB-Refining further mitigates channel aging over long data frames by employing reliably detected data symbols as virtual pilots for periodic channel re-estimation. Simulation results demonstrate rapid convergence and improved channel NMSE and SER compared with EKF-EP, EKF-ZF, and JED baselines considered in this work. These results also show that tracking the evolving PN process across the pilot and data phases is important for maintaining detection performance when PN varies over time. Future work will investigate online estimation of PN innovation statistics, wideband frequency-selective channels with PN, and generative priors to address severe and non-stationary PN environments.

\appendix
\subsection{Proof of Theorem \ref{thm:collapsedVM}}
\label{append:B}
To find the single von Mises distribution $q(\theta) = \mc{VM}(\theta;\mu,\kappa)$ that best approximates the mixture posterior $p(\theta) = \sum_{k=1}^{N}\alpha_k \mc{VM}(\theta;\eta_k)$ in the minimum KL-divergence sense, we match the first trigonometric moment $\mathbb{E}[\e^{\ji\theta}]$.

By the linearity of expectation over the mixture components,
\begin{align}
\mathbb{E}_{p}\!\left[\e^{\ji\theta}\right]
&= \sum_{k=1}^{N}\alpha_k \int_{-\pi}^{\pi} \mc{VM}(\theta;\eta_k)\,\e^{\ji\theta}\,d\theta \nonumber\\
&= \sum_{k=1}^{N}\alpha_k \mathbb{E}_{\mc{VM}(\eta_k)}\!\left[\e^{\ji\theta}\right].
\end{align}
Using the first moment identity for a von Mises component, $\mathbb{E}_{\mc{VM}(\eta_k)}\!\left[\e^{\ji\theta}\right] = J(|\eta_k|)\,\e^{\ji\angle\eta_k}$, we obtain:
\begin{equation} \label{eq:mixVM}
\mathbb{E}_{p}\!\left[\e^{\ji\theta}\right] = \sum_{k=1}^{N} \alpha_k J(|\eta_k|) \e^{\ji\angle\eta_k}.
\end{equation}
Equating this to the first trigonometric moment of the target distribution $\mc{VM}(\theta;\mu,\kappa)$, given by $\mathbb{E}_{q}[\e^{\ji \theta}] = J(\kappa)\e^{\ji \mu}$, yields:
\begin{equation} 
J(\kappa)\e^{\ji \mu} = \bigg|\sum_{k=1}^{N} \alpha_k J(|\eta_k|) \e^{\ji\angle\eta_k}\bigg| \e^{ \ji\angle\left(\sum_{k=1}^{N} \alpha_k J(|\eta_k|) \e^{\ji\angle\eta_k}\right)}.
\end{equation}
Separating the phase and modulus gives the expressions for $\mu$ and $\kappa$ in \eqref{eq:mean:theta} and \eqref{eq:con:theta}, completing the proof.

\bibliographystyle{IEEEtran}
\bibliography{JED_refs}
\end{document}